\documentclass[aps,prc,twocolumn,nofootinbib,superscriptaddress,amsmath,amssymb]{revtex4-1}

\usepackage{amsmath}
\usepackage{amssymb}
\usepackage{graphicx}
\usepackage{bm}
\usepackage{hyperref}

\begin{document}

\title{{\it Ab initio} constraints on thermal effects of the nuclear equation of state}

\author{Arianna Carbone}
\email{acarbone@ectstar.eu}
\affiliation{European Centre for Theoretical Studies in Nuclear Physics and Related Areas (ECT*)
and Fondazione Bruno Kessler, Strada delle Tabarelle 286, I-38123 Villazzano (TN), Italy}
\author{Achim Schwenk}
\email{schwenk@physik.tu-darmstadt.de}
\affiliation{Institut f\"ur Kernphysik, 
Technische Universit\"at Darmstadt, 
64289 Darmstadt, Germany}
\affiliation{ExtreMe Matter Institute EMMI, 
GSI Helmholtzzentrum f\"ur Schwerionenforschung GmbH, 
64291 Darmstadt, Germany}
\affiliation{Max-Planck-Institut f\"ur Kernphysik, 
Saupfercheckweg 1, 
69117 Heidelberg, 
Germany}

\begin{abstract}
We exploit the many-body self-consistent Green's function method to analyze
finite-temperature properties of infinite nuclear matter and to explore the 
behavior of the thermal index used to simulate thermal effects in equations of state
for astrophysical applications. We show how the thermal index is both density 
and temperature dependent, unlike often considered, and we provide an error 
estimate based on our {\it ab initio} calculations. The inclusion of three-body forces
is found to be critical for the density dependence of the thermal index. We also
compare our results to a parametrization in terms of the density dependence
of the nucleon effective mass. Our findings point to possible shortcomings of predictions 
made for the gravitational-wave signal from neutron-star merger simulations
with a constant thermal index.
\end{abstract}

\maketitle

\section{Introduction}

The nuclear equation of state (EOS) is an essential relation between pressure, density, and temperature, which provides proper closure for the solution of the equations of motion in relativistic hydrodynamics~\cite{Duez2018}. These equations are numerically implemented to simulate, among others, the merger of neutron stars~\cite{Baiotti2017}. The outcome of these astrophysical simulations provides insights to the gravitational-wave signal produced by such events or the modeling of short gamma-ray bursts arising from the merging~\cite{Shibata2005,Anderson2008,Baiotti2008,Bernuzzi2012,Bauswein2012}. During these phenomena, the temperature of matter can rise to extremes, $T \sim 100$~MeV. In order to obtain reliable results from these simulations, it is then mandatory to consider EOSs that correctly describe the thermal effects of dense matter~\cite{Oechslin2007}. 

In view of the recent detection of a gravitational-wave signal from the merger of two neutron stars~\cite{gw170817}, it is timely for theoretical studies to provide accurate results to meet the needs for correct interpretation of observational outcomes. Apart from the complexity in solving the equations of relativistic hydrodynamics, the modeling of mergers  needs to be improved regarding the knowledge of the nuclear matter EOS~\cite{Hebe13ApJ,Oertel2017,Baym2018,Burgio2018,Raithel:2019gws}. This is caused by the challenges in understanding the properties of nuclear interactions and dense matter, which are governed by the theory of quantum chromodynamics (QCD). At nuclear densities, the relevant degrees of freedom are nucleons and pions (and possible $\Delta$ isobars), and most studies of nuclear structure, reactions and matter employ systematic many-body methods with internucleon interactions (see, e.g., Refs.~\cite{Heiselberg2000,Hebeler2015,Hjorth-Jensen2017}) or by means of density functional theory (see, e.g., Refs.~\cite{Bender2003,Kortelainen2010}). While the former are based on nuclear forces constructed to reproduce  nucleon-nucleon scattering and properties of light nuclei~\cite{Wiringa1995,Epelbaum2009,Mach11PR,Hamm13RMP}, the latter are fit to selected nuclei and often nuclear matter~\cite{Stone2007}. In this work, we use modern nuclear forces derived from chiral effective field theory (EFT)~\cite{Epelbaum2009,Mach11PR,Hamm13RMP} and solve the nuclear many-body problem by means of a nonperturbative many-body approach, the self-consistent Green's function (SCGF) method~\cite{Dickhoff2004,Barbieri2017}. Nuclear matter at finite temperature has also been studied based on chiral low-momentum interactions within many-body perturbation theory (see, e.g., Refs.~\cite{Tolo08nmatt,Well14nmtherm}). Moreover, while nuclear matter at $T=0$ has been the target of many works, there have been a range of more recent advances based on chiral interactions, including Refs.~\cite{Hebeler2010,Hebeler2011,Tews:2012fj,Gezerlis:2013ipa,Holt:2013fwa,Hagen:2013yba,Coraggio:2014nva,Carbone2014,Lynn:2015jua,Drischler:2016djf,Holt:2016pjb,Drischler:2017wtt,Bombaci2018}.

Being derived from a low-energy expansion of QCD, chiral EFT interactions are organized using a power counting scheme in terms of powers of a low-momentum scale over the breakdown scale~\cite{Weinberg1990,Weinberg1991}. The high-energy details, which are not resolved at low energies, are then encoded in the strength of short-range contact interactions. The remaining contributions are given by pion exchanges, and can include also $\Delta$ degrees of freedom. In this study, we will consider two-nucleon $(2N)$ interactions up to third (N2LO)~\cite{Ekstroem2015} or fourth order (N3LO)~\cite{Entem2003} in the chiral expansion, while the three-nucleon $(3N)$ forces will be considered at N2LO~\cite{VanKolck1994,Epelbaum2002}. We use different chiral two- and three-nucleon interactions within the finite-temperature SCGF method to study the properties of infinite nuclear matter. This approach is particularly suited because it is implemented directly at finite temperature, and so it provides the full thermodynamical properties of nuclear matter. The method is nonperturbative providing a self-energy, which resums all particle-particle and hole-hole diagrams, leading to a fully dressed definition of the single-particle Green's function, from which the nuclear matter bulk properties are accessed~\cite{Frick2003,Rios2009,Soma2009}.

Presenting first-principle calculations of the nuclear-matter energy and pressure at finite temperature, we demonstrate the shortcoming of taking an ideal gas to model the thermal contributions to the EOS in astrophysical simulations~\cite{Shibata2005,Oechslin2007,Baiotti2008,Hotokezaka2011,Bauswein2012,Rezzolla2016,Endrizzi2018}. With our {\it ab initio} calculations, we explore how the thermal index $\Gamma_{\rm th}$, which describes the thermal effects of the EOS, in fact depends on density (and to a lesser extent temperature) in a way that has not been explored in simulations. This inevitably leads us to the conclusion that a constant $\Gamma_{\rm th}$ value provides only a crude approximation to the EOS of dense matter and may lead to shortcomings in the predictions when simulations involve higher temperatures. The effects of using a constant versus varying index in the EOS has been analyzed in studies of the gravitational-wave signal from merging neutron stars~\cite{Bauswein2010}. Furthermore, this had been investigated in the propagation of fast magnetosonic shocks in relativistic fluids~\cite{Mignone2007}. Recent studies within Fermi liquid theory and mean-field approaches, have also highlighted the non-constant behavior of the thermal  index~\cite{Constantinou2015,Lattimer2016}. We point out that several studies of neutron star mergers already include fully finite-temperature EOS (see, e.g., Refs.~\cite{Bovard:2017mvn,Radice2018a,Radice2018b,Perego2019}), however based on the mean-field approximation.

The paper is organized as follows. In Sec.~\ref{sec:formalism} we discuss the many-body SCGF formalism used to study nuclear matter. We then follow with our results in Sec.~\ref{sec:results}, which is divided into three parts. First, we present results for the thermal energy, thermal pressure, and thermal index employing one particular chiral two- and three-nucleon interaction and analyze the behavior with density and temperature. Second, we provide an error estimate for the thermal effects using several chiral Hamiltonians. Then, we characterize the thermal index through the density dependence of the nucleon effective mass, which we obtain directly from our SCGF calculations as well. Finally, we conclude in Sec.~\ref{sec:conclusions} and give an outlook.

\begin{figure*}
\centering
\includegraphics[width=0.45\textwidth]{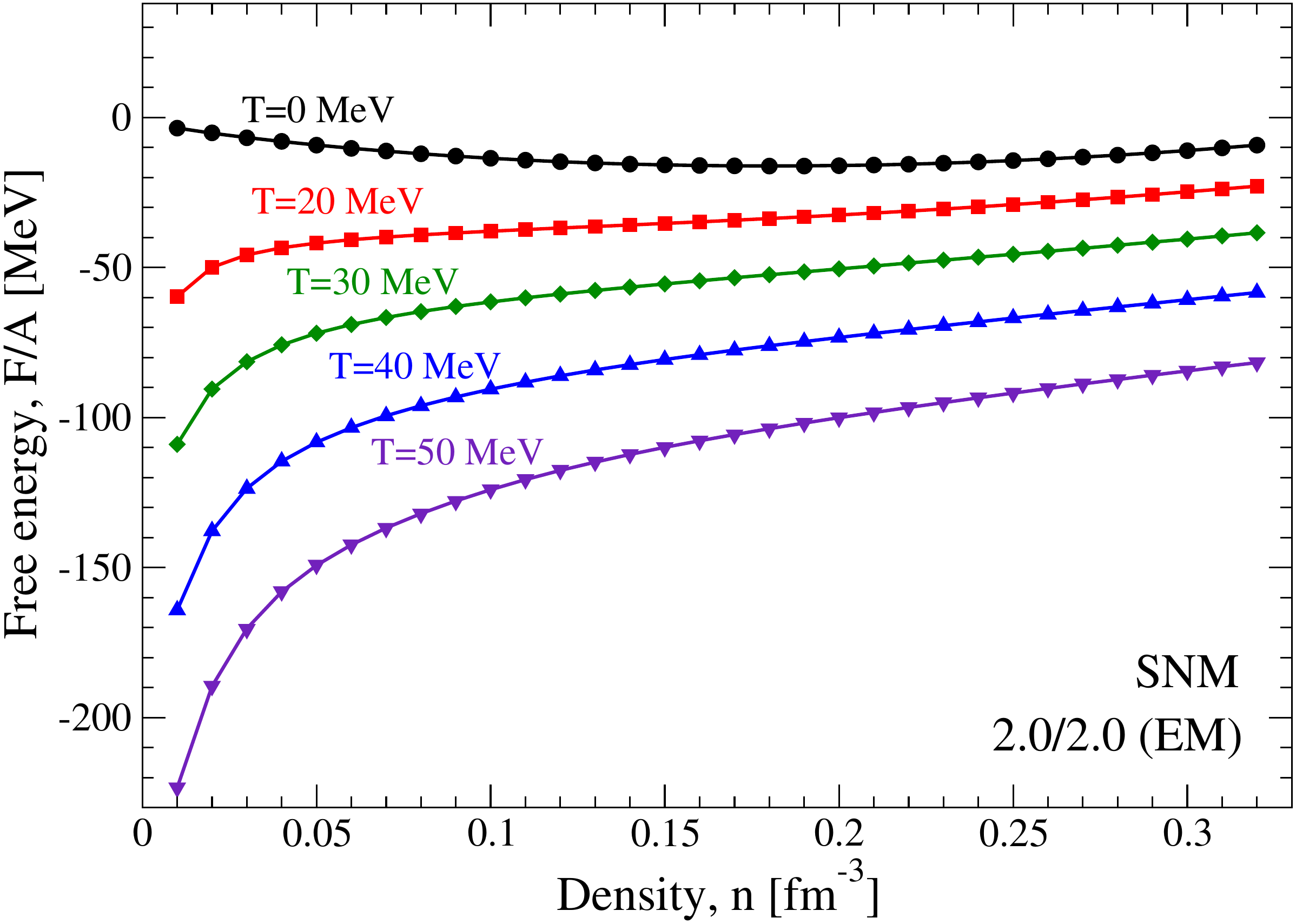}
\hspace*{5mm}
\includegraphics[width=0.45\textwidth]{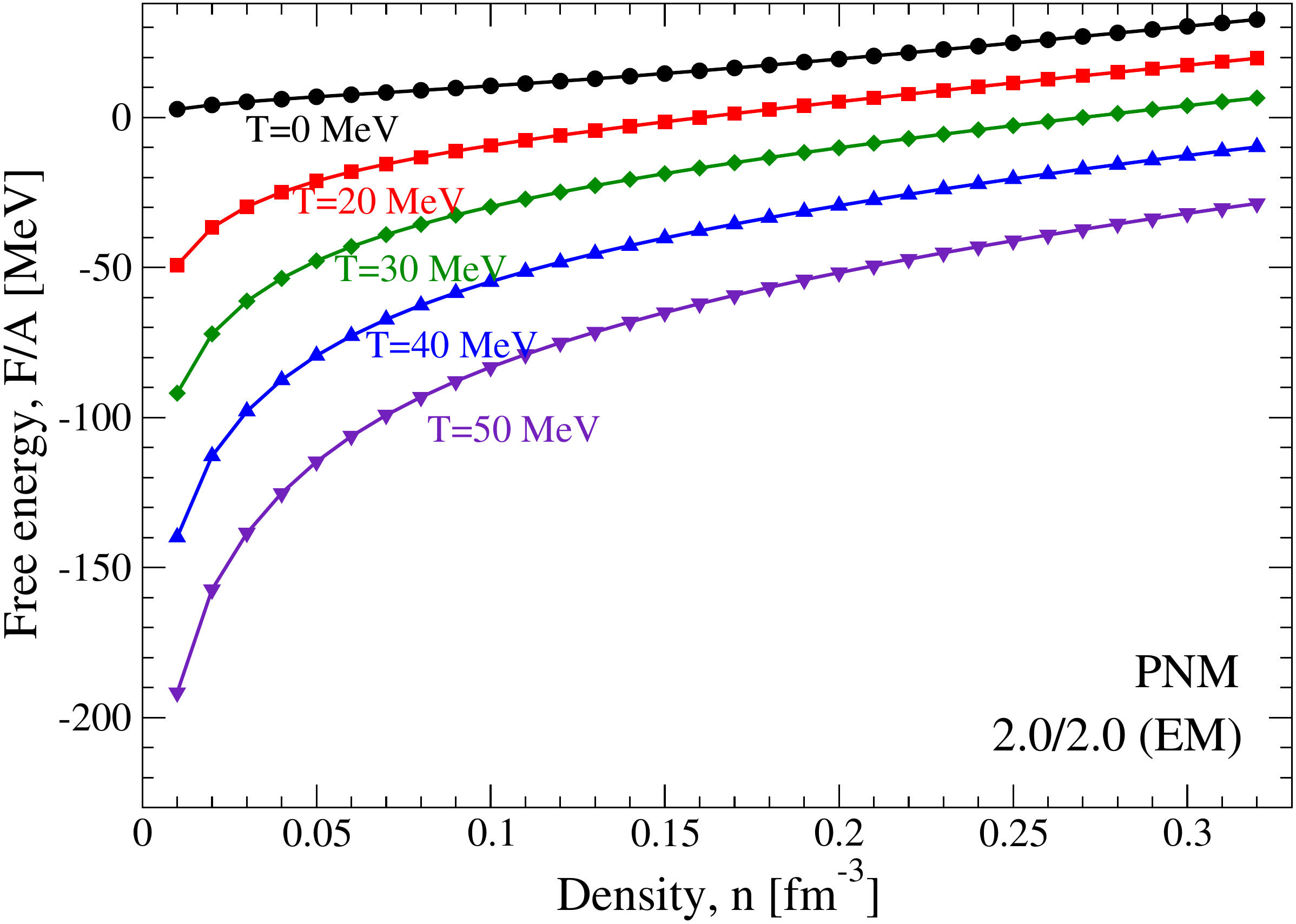}
\caption{Energy per nucleon (at $T=0$, black dots/lines) and free energy per nucleon
for $T=20, 30, 40$, and 50~MeV as function of density for SNM (left) and PNM (right), 
obtained from SCGF calculations with the 2.0/2.0 (EM) chiral two-
and three-nucleon interactions. Symbols correspond
to calculated data, solid lines represent the fits of the free-energy (see text for details).}
\label{fig:free_ener}
\end{figure*}

\section{Formalism: Finite-temperature properties of nuclear matter}
\label{sec:formalism}

To access nuclear matter at finite temperature we use the SCGF approach. As already introduced, this method fits well for our present study because it is implemented directly at finite temperature~\cite{Frick2003,Rios2009}. Furthermore, it in principle provides a thermodynamically consistent description of matter, meaning that physical properties calculated microscopically or through macroscopical (thermodynamic) relations should equal one another~\cite{Baym1961,Baym1962}. The method is based on the definition of a self-energy, which resums an infinite series of particle-particle and hole-hole diagrams, also known as ladder approximation for the self-energy. This self-energy is then employed to construct a fully dressed single-particle propagator, the Green's function $G$, from the free Green's function $G_0$, via solution of the Dyson equation:
\begin{equation}
G({\bf p},\omega)=G_{0}({\bf p},\omega)+G_{0}({\bf p},\omega)\Sigma^\star({\bf p},\omega)G({\bf p},\omega)\,,
\label{eq:dyson}
\end{equation} 
where ${\bf p}$ and $\omega$ are the single-particle momentum and energy, and $\Sigma^\star$ is the irreducible self-energy. The imaginary part of the Green's function yields the spectral function, which is the central quantity used to calculate both microscopic as well as bulk properties of the many-body system. The spectral function describes the probability of adding or removing a particle with momentum ${\bf p}$ and energy $\omega$ to or from the many-body system.

In the past years, the SCGF approach has been extended to include three-body forces, i.e., to start from a Hamiltonian $H=T_{\rm kin}+V$ where the interacting part includes two- and three-nucleon interactions, $V=V_{2N}+V_{3N}$~\cite{Carbone2013II}. Through knowledge of the spectral function, one can access the energy per nucleon $E/A$ of the system employing the Galitskii-Migdal-Koltun sum rule~\cite{Galitskii1958,Koltun1974}:
\begin{equation}
\frac{E}{A}=\frac{\nu}{n}\int\frac{{\rm d}{\bf p}}{(2\pi)^3}\int\frac{{\rm d}\omega}{2\pi} \, \frac{1}{2}\Big[\frac{p^2}{2m}+\omega\Big]\mathcal{A}({\bf p},\omega)f(\omega)-\frac{1}{2}\langle V_{3N} \rangle\,,
\label{eq:energy}
\end{equation}
where $\nu$ is the degeneracy of the system, $n$ the number density, $\mathcal{A}({\bf p},\omega)$ is the spectral function, $f(\omega)$ is the Fermi-Dirac distribution function. $\langle V_{3N} \rangle$ is the expectation value of the three-body operator; at present we only approximate this quantity with its first-order term but calculate it employing fully dressed propagators~\cite{Carbone2014}. The free energy $F$ is then obtained from
\begin{equation}
\frac{F}{A}=\frac{E}{A}-T \,\frac{S}{A}\,. 
\label{eq:free}
\end{equation} 
The entropy $S/A=-\partial\Omega/\partial T|_\mu$ is evaluated following the Luttinger-Ward formalism, which demonstrates that it is possible to define the grand-canonical potential $\Omega$ in terms of the Green's function $G$~\cite{Luttinger1960}. Detailed description of the calculation of the entropy can be found in Ref.~\cite{riosphd}. From the free energy one can then directly access the pressure $P$ via
\begin{equation}
P=n^2 \, \frac{\partial F/A}{\partial n}\,.
\label{eq:pressmacro}
\end{equation}
As we stated above, the SCGF method is a thermodynamically consistent approach, so in principle we could also calculate the pressure starting from the microscopic chemical potential, $\tilde\mu$, via
\begin{equation}
\tilde P = n(\tilde\mu-\frac{F}{A})\,,
\label{eq:pressmicro}
\end{equation}
where $\tilde\mu$ is obtained inverting the density sum rule,
\begin{equation}
n=\nu \int\frac{{\rm d}{\bf p}}{(2\pi)^3}\int\frac{\rm d\omega}{2\pi}\mathcal{A}({\bf p},\omega)f(\omega)\,,
\label{eq:mumicro}
\end{equation}
with $f(\omega)=[1+e^{(\omega-\tilde\mu)/T}]^{-1}$. 

In view of the thermodynamical consistency of this approach, the equality $P=\tilde P$ should hold, up to numerical errors. We have tested wether our calculations reproduce this equality and have encountered some discrepancies which are density dependent (see Ref.~\cite{Carbone2018} for details). In fact these differences depend on the strength of the three-body forces. Calculations with only two-nucleon interactions prove the equality true. We consider the error as coming from the approximation we perform on the $\langle V_{3N} \rangle$ expectation value in the evaluation of the energy sum rule, Eq.~\eqref{eq:energy}. Improvements that go beyond this are work in progress. However, given that the thermal index $\Gamma_{\rm th}$ depends on differences of pressures, $P(T)-P(T=0)$, at a fixed density, and that these discrepancies are mostly density dependent and only mildly temperature dependent, we can rely on the use of either $P$ or $\tilde P$ to obtain $\Gamma_{\rm th}$. A detailed analysis of this will be discussed in future work~\cite{Carboneunpub}, but overall uncertainties are small compared to the combined SCGF and interaction uncertainties.

\begin{figure*}
\centering
\includegraphics[width=0.45\textwidth]{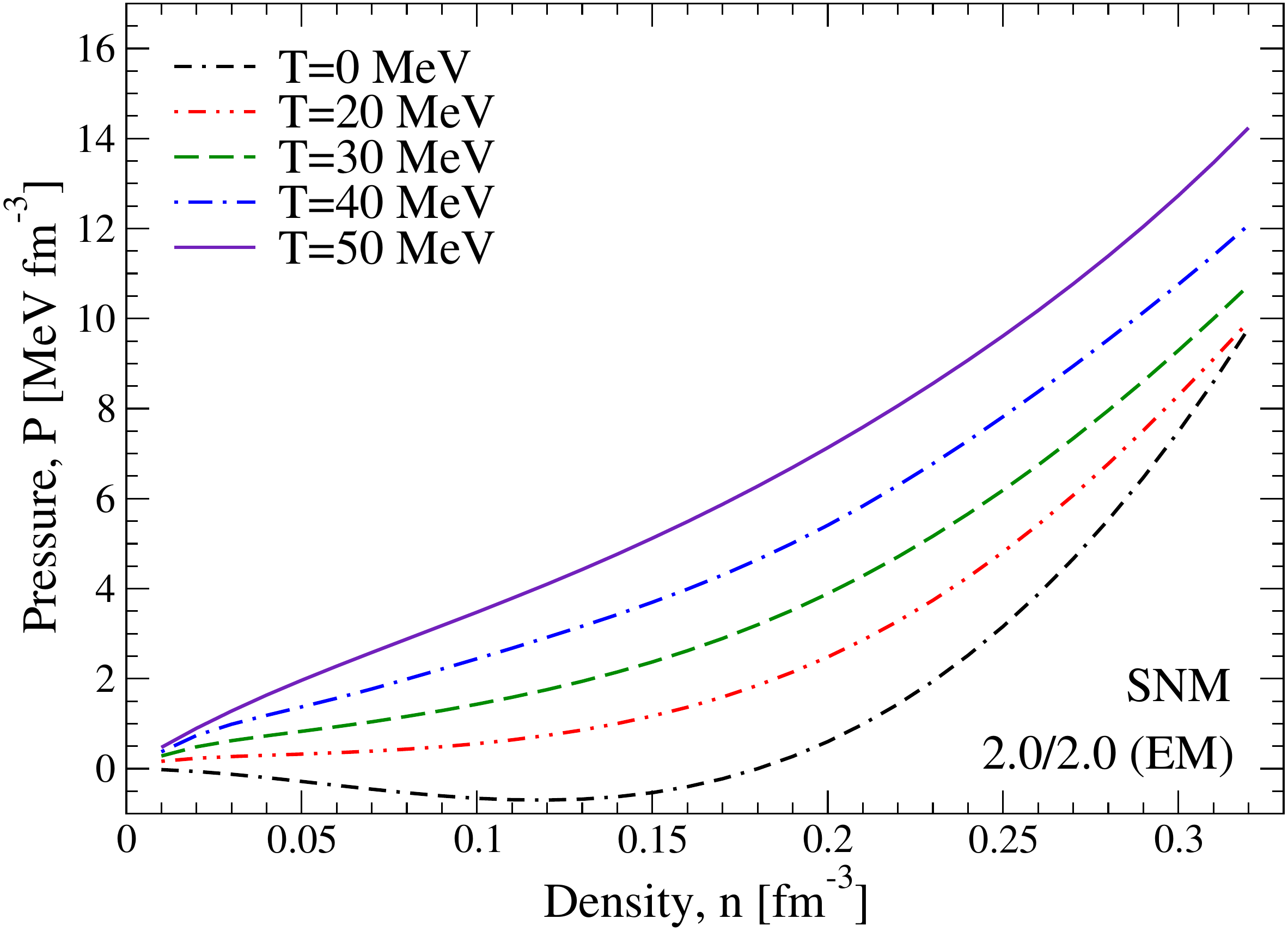}
\hspace*{5mm}
\includegraphics[width=0.45\textwidth]{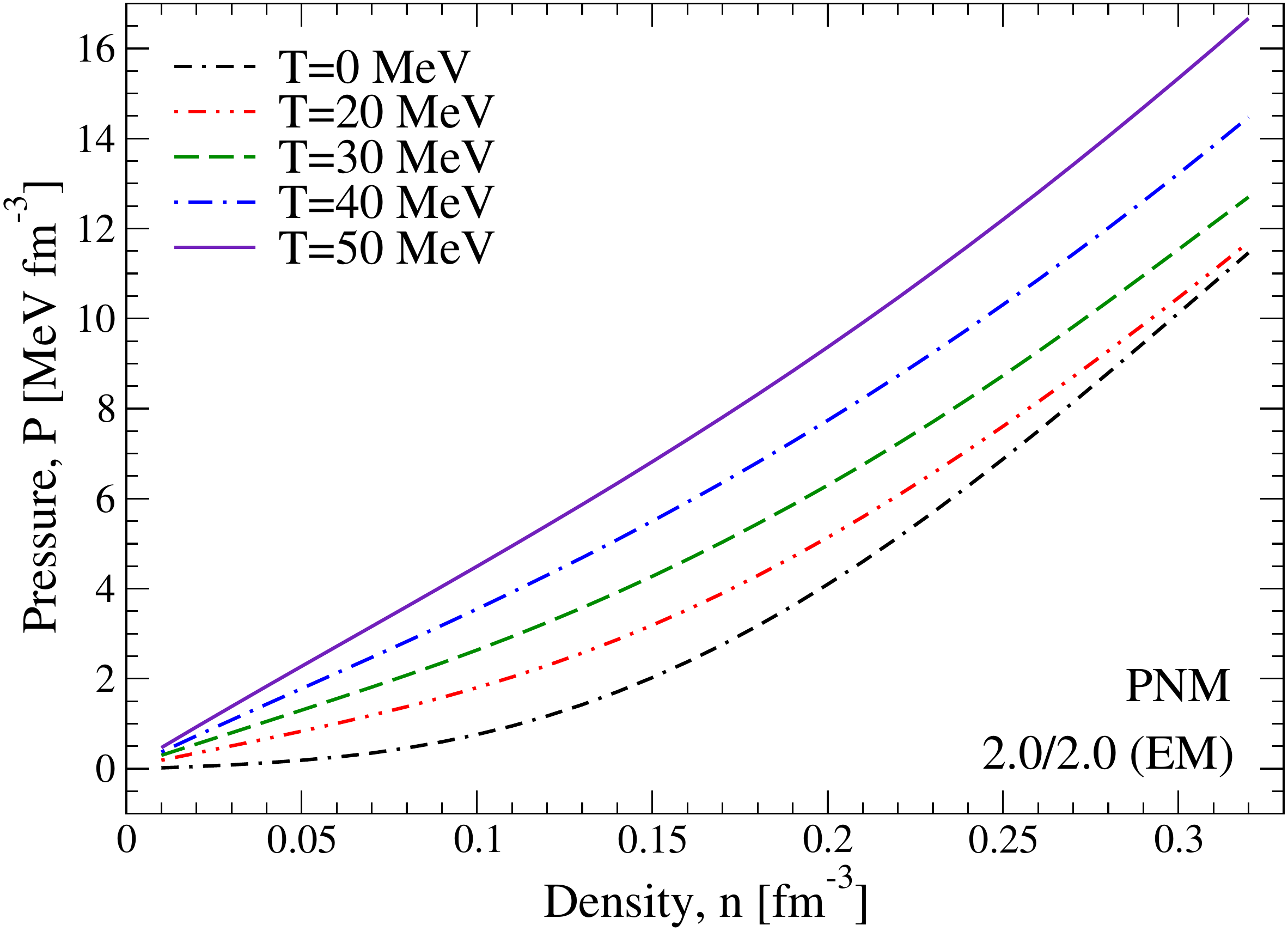}
\caption{Pressure for $T=0, 20, 30, 40$, and 50~MeV as function of density for SNM (left) 
and PNM (right), obtained from SCGF calculations with the 2.0/2.0 (EM) chiral two-
and three-nucleon interactions.}
\label{fig:press}
\end{figure*}

In this work, we use the pressure obtained as a derivative of the free energy, Eq.~\eqref{eq:pressmacro}. In order to perform the density derivative, we fit the calculated SCGF results with a similar function to that presented in Ref.~\cite{Drischler2016} and extended to finite temperature in Ref.~\cite{Carbone2018}: 
\begin{equation}
\frac{F}{A}(n,T)=a(T)+\sum_{\nu=2}^{\nu_{\rm max}}a_\nu(T)\left(\frac{n}{n_0}\right)^{\nu/3}\,,
\label{eq:freefit}
\end{equation}
where $\nu_{\rm max} $=6-10 is used depending on the two- and three-nucleon interactions, and $n_0$ is a fiducial density taken to be the saturation density $0.16$~fm$^{-3}$. In Fig.~\ref{fig:free_ener} we plot the energy per nucleon at $T=0$ and the free energy per nucleon for four different temperatures, $T=20,30,40$, and 50~MeV, for  symmetric nuclear matter (SNM) and pure neutron matter (PNM). The results shown are based on the 2.0/2.0 (EM) chiral two- and three-nucleon interactions (see Ref.~\cite{Hebeler2011} for details). The points give the SCGF results obtained from Eqs.~\eqref{eq:energy} and~\eqref{eq:free}. The goodness of the fit given by Eq.~\eqref{eq:freefit} can be appreciated through the solid lines. Correspondingly, the pressure obtained from Eq.~\eqref{eq:pressmacro} is presented in Fig.~\ref{fig:press} for SNM and PNM.

\section{Results: Thermal effects of the nuclear equation of state}
\label{sec:results}

In astrophysical applications, the thermal contributions to the EOS are
often modelled following an ideal gas \cite{Rezzolla_book:2013}. In this
case, the pressure as a function of the number density $n$ and the energy density $E/V$ 
is expressed in terms of an adiabatic index $\Gamma$:
\begin{equation}
P_{\rm th}(n,E)= \frac{E_{\rm th}}{V}(\Gamma-1)\,,
\label{eq:idealgas}
\end{equation}
with the volume $V=A/n$.
Unlike a polytropic EOS, which describes isentropic processes, i.e.,
adiabatic and reversible, the pressure given by Eq.~(\ref{eq:idealgas})
allows non-isentropic irreversible transformations, such as ``shock
heating'', by means of which kinetic energy can be transformed into
internal energy, thus increasing the temperature of the system. These
shocks could be produced, as an example, during the merging of two
neutron stars. For this reason, one usually writes the pressure as
\begin{equation}
P=P_{0}+P_{\rm th}\,,
\label{eq:spliteos}
\end{equation}
where only the thermal part of the pressure, $P_{\rm th}$, is modelled
employing Eq.~(\ref{eq:idealgas}), while the cold part, $P_{0}=P(T=0)$,
can also be a polytropic function or a microscopic EOS at $T=0$. We note
that Eq.~(\ref{eq:idealgas}) has at least two shortcomings. First, the
adiabatic index $\Gamma$ in Eq.~(\ref{eq:idealgas}) is often taken as a
constant because it is associated with the ratio of the specific heats in
the fluid \cite{Rezzolla_book:2013}. Second, Eq.~(\ref{eq:idealgas}) has
been shown to be incompatible with relativistic kinetic theory for
arbitrary values of $\Gamma$, whose value should instead depend on the
quantity $P/n$ in order to fulfill the so-called Taub's
inequality~\cite{Taub1948, Rezzolla_book:2013}. In view of these
considerations, it is better to express the thermal index in terms of the thermal pressure and thermal energy as:
\begin{equation}
\Gamma_{\rm th}=1+\frac{P_{\rm th}}{E_{\rm th}/V}\,,
\label{eq:thindex}
\end{equation}
where $P_{\rm th}=P(T)-P_0$ and $E_{\rm th}=E(T)-E(T=0)$ are the thermal
pressure and thermal energy, respectively. It should be noted that this
formulation is consistent with relativistic kinetic theory and Taub's
inequality~\cite{Mignone2007}. Given that we have access to the thermal
pressure and thermal energy from our SCGF calculations, we can then
investigate the behavior of the thermal index, as diagnostic of the
thermal effects, based on modern two- and three-nucleon interactions.  Using the thermal index $\Gamma_{\rm th}$
obtained in this way to replace the constant adiabatic $\Gamma$ index in
Eq.~(\ref{eq:idealgas}) will provide a more accurate representation of the
thermal effects of nuclear matter.

\subsection{Thermal energy, thermal pressure, and thermal index from {\it ab initio} calculations}
\begin{figure*}
\centering
\includegraphics[width=0.45\textwidth]{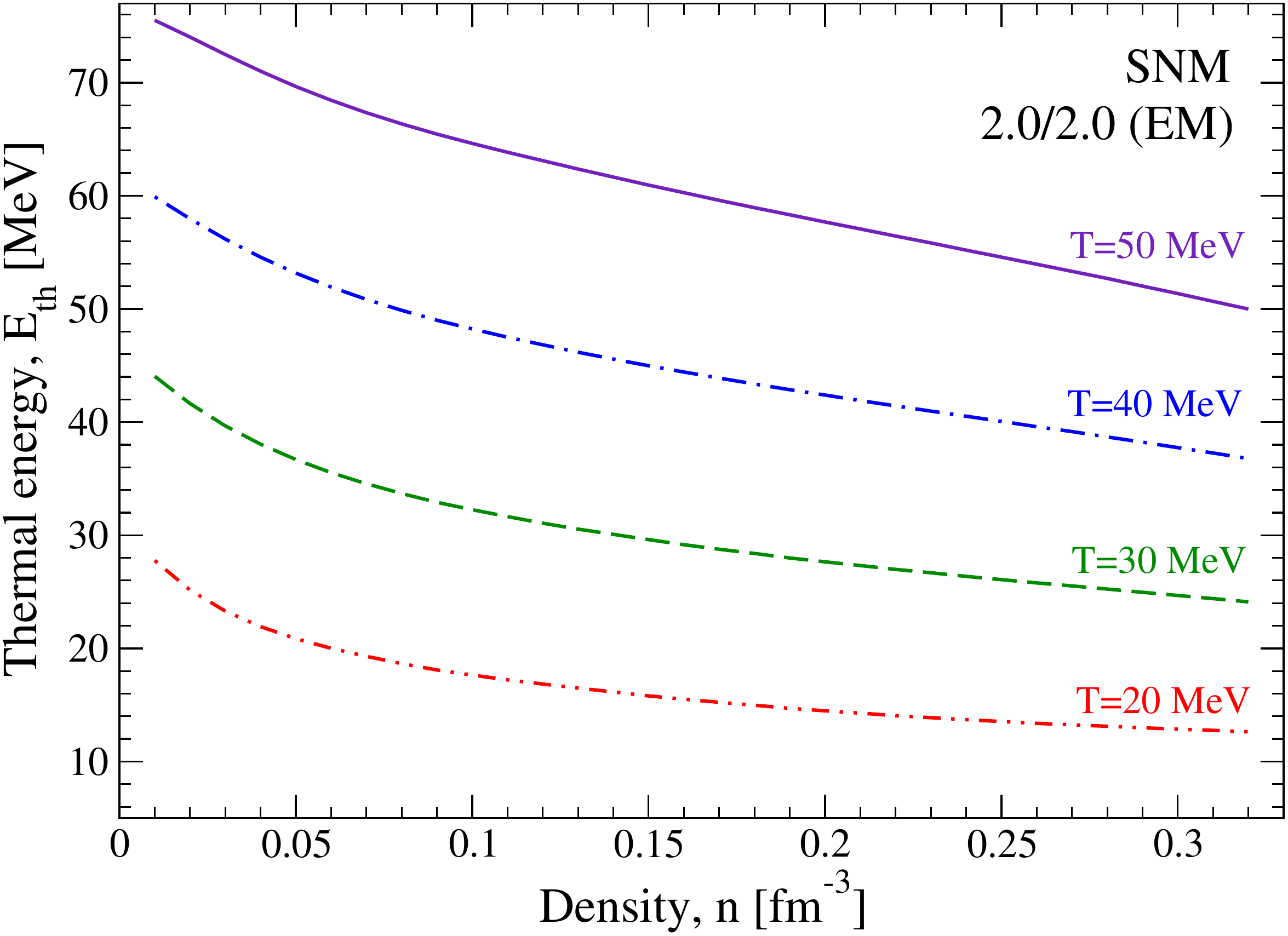}
\hspace*{5mm}
\includegraphics[width=0.45\textwidth]{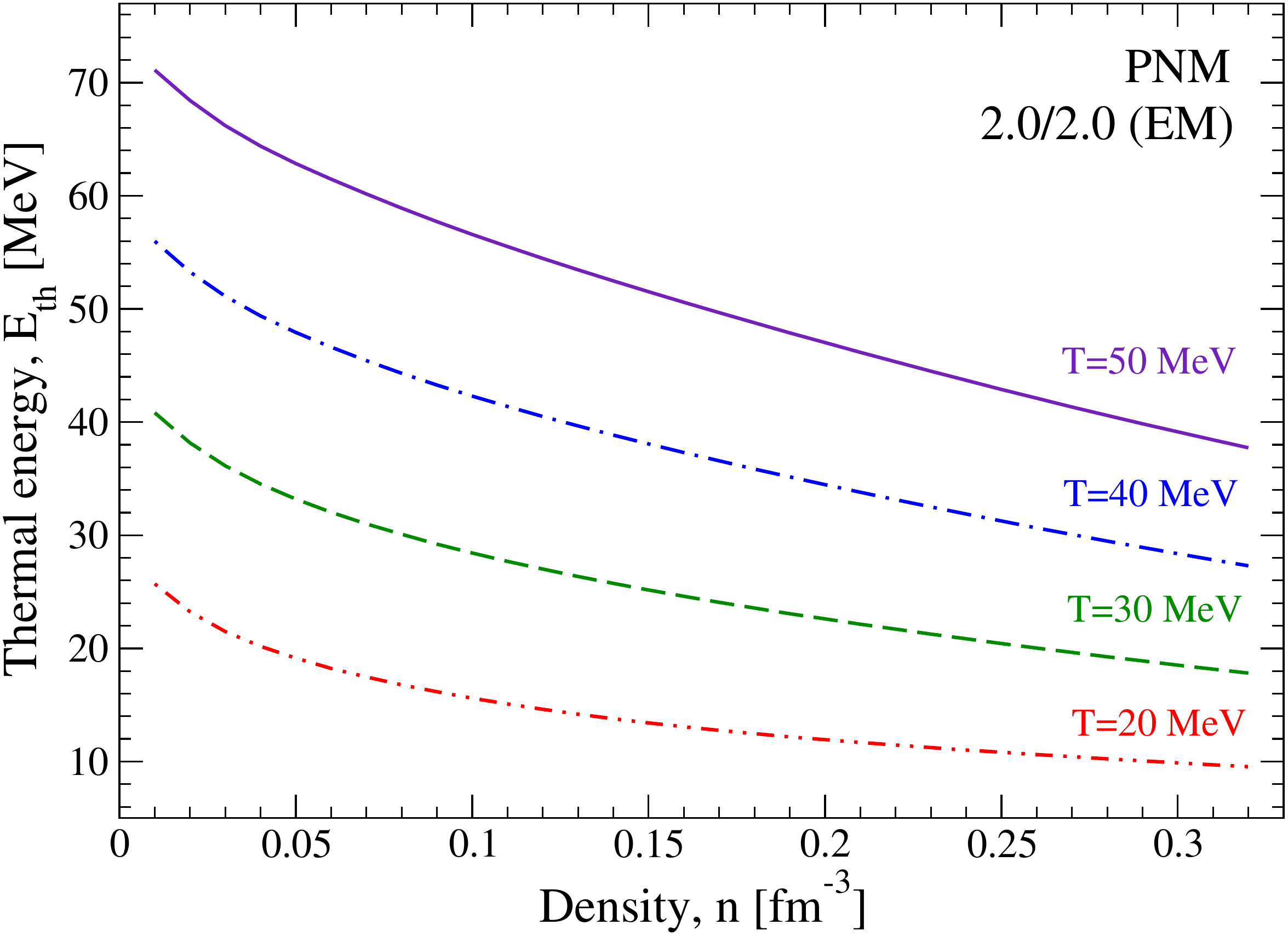}
\includegraphics[width=0.45\textwidth]{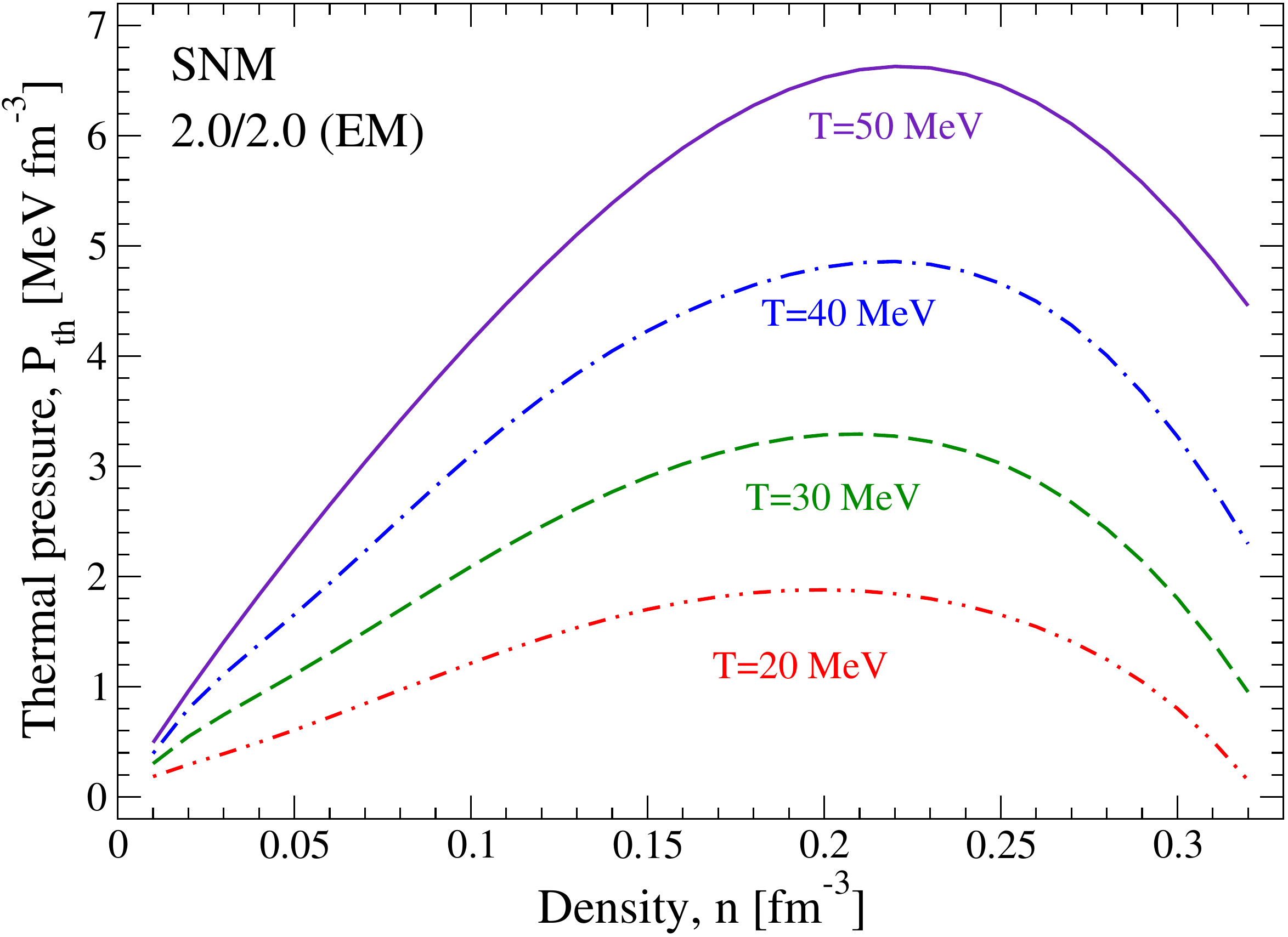}
\hspace*{5mm}
\includegraphics[width=0.45\textwidth]{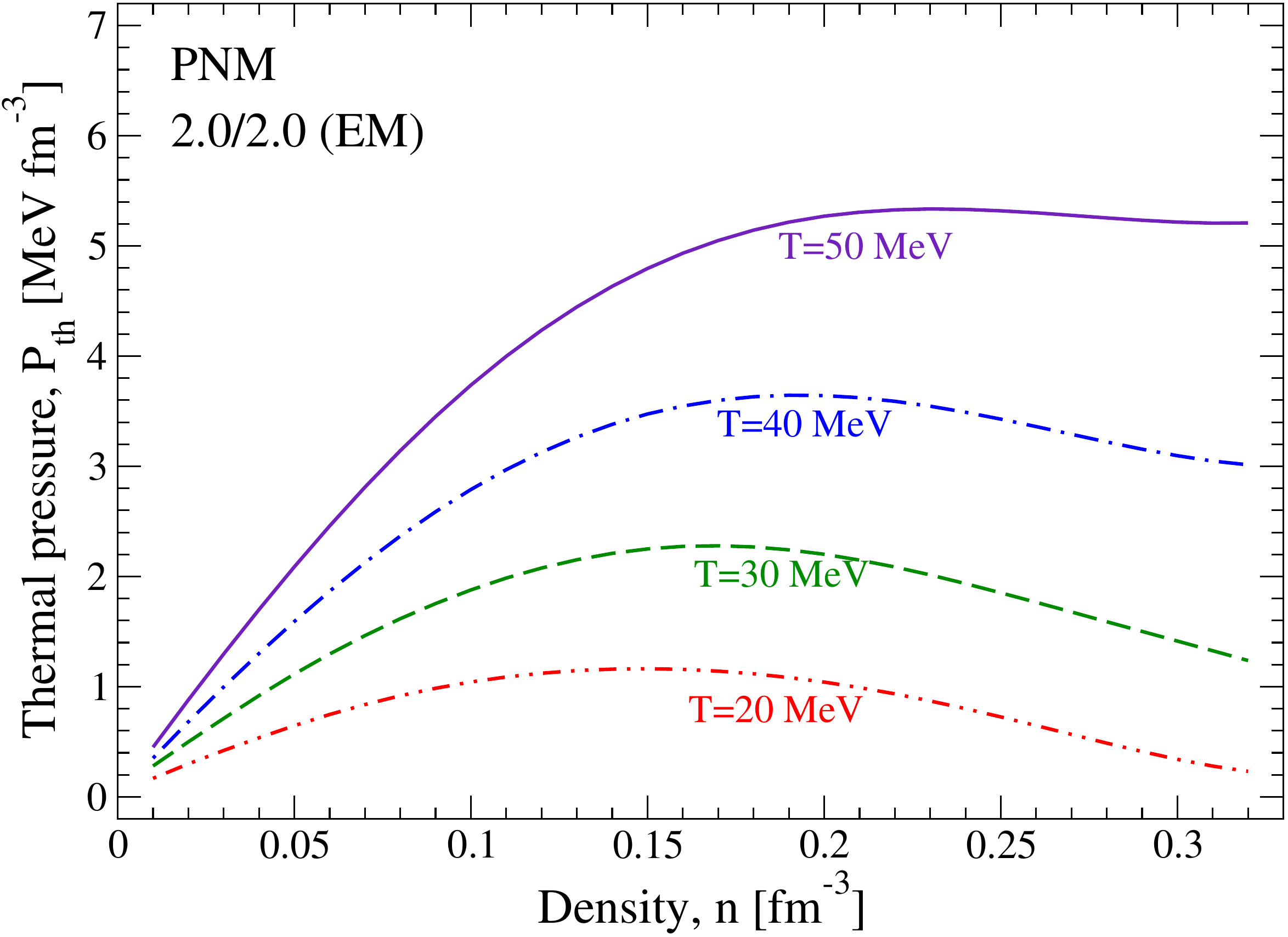}
\includegraphics[width=0.45\textwidth]{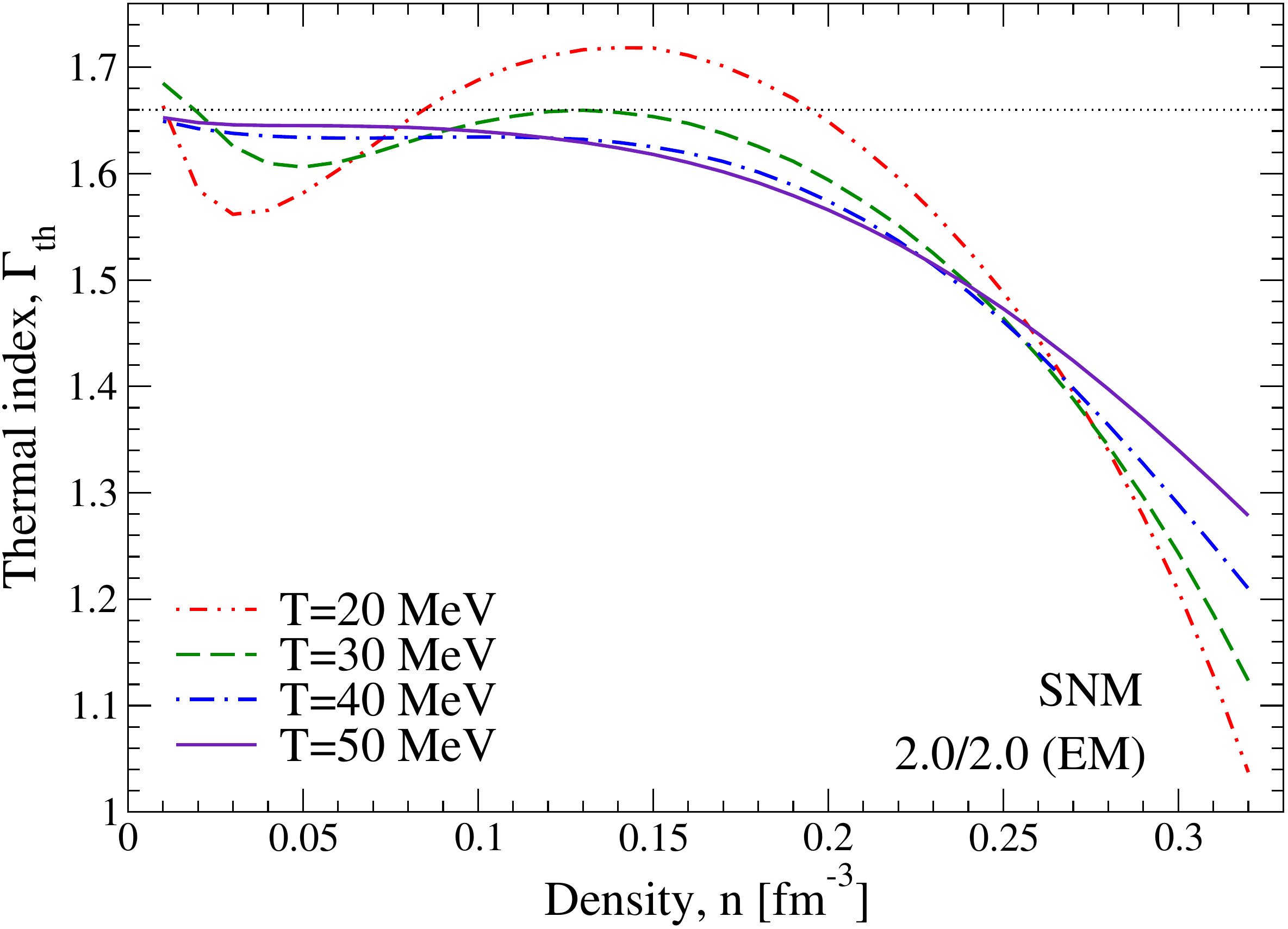}
\hspace*{5mm}
\includegraphics[width=0.45\textwidth]{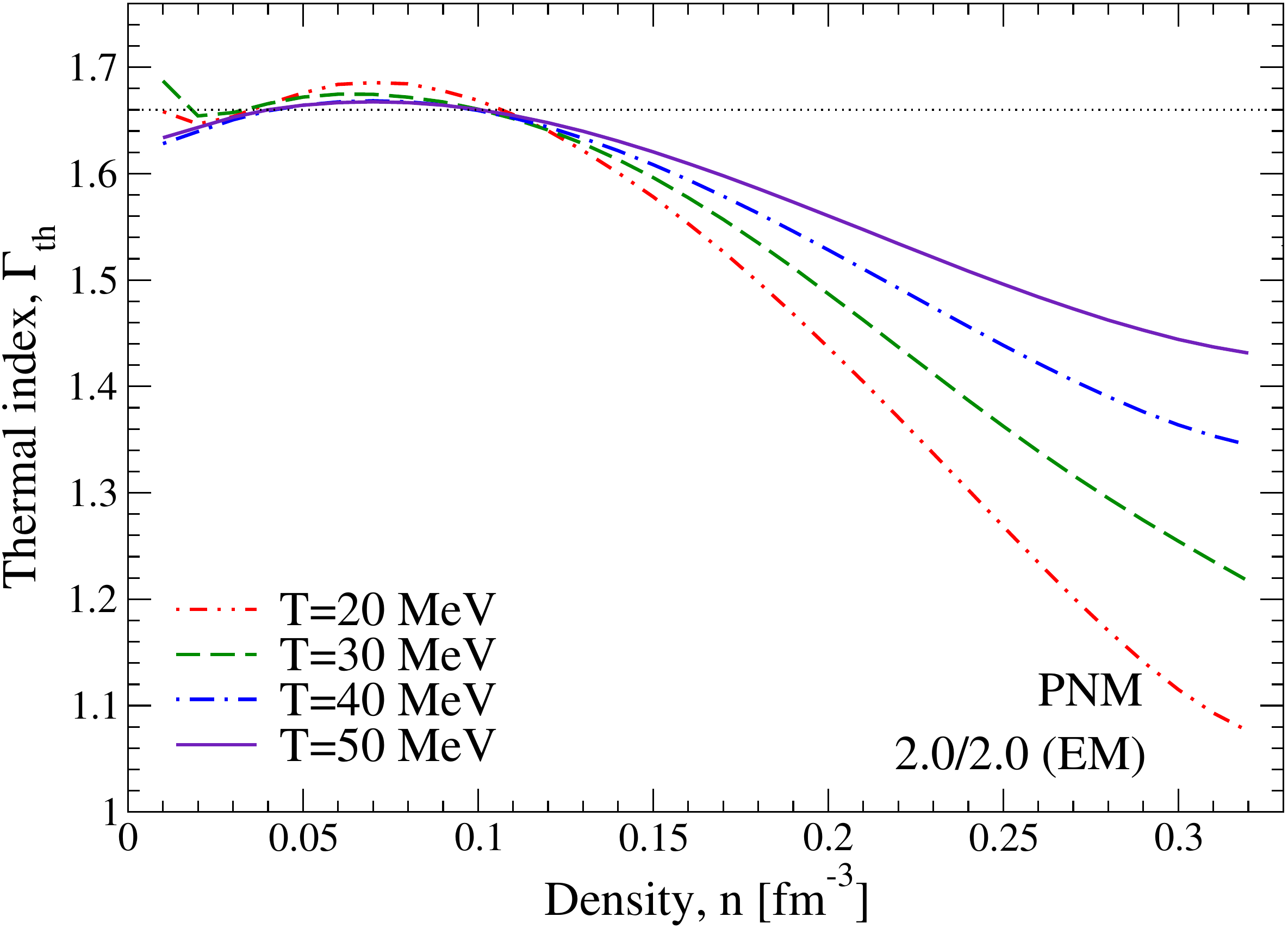}
\caption{Thermal energy (first row), thermal pressure (second row), and thermal index 
(third row) for $T=20, 30, 40$, and 50~MeV as function of density, for SNM (left panels) 
and PNM (right panels), obtained from SCGF calculations with the 2.0/2.0 (EM) chiral two-
and three-nucleon interactions.}
\label{fig:therm_srg20}
\end{figure*}

We present in Fig.~\ref{fig:therm_srg20} the thermal energy, thermal pressure, and thermal index for SNM (left panels) and PNM (right panels). We employ the same chiral two- and three-nucleon interactions as the one used in Figs.~\ref{fig:free_ener} and~\ref{fig:press}. The thermal quantities are presented for four different temperatures, $T=20,30,40,$ and 50~MeV.

The thermal energy decreases with increasing density for all studied temperatures and for both SNM and PNM. At low densities, up to $n=A/V=0.05$~fm$^{-3},$ the decrease appears steeper, especially in the case of PNM. This is because at constant temperature, matter is more nondegenerate at low density leading to stronger thermal effects in the low density regime, while as the density increases, the difference between the finite-temperature energy and its value at zero temperature becomes smaller. For low temperatures, $T=20$~MeV, the thermal energy becomes very weakly density dependent already around twice saturation density. However, as $T$ increases, the thermal energy decreases up to twice saturation density, which is the limit in densities we consider for our calculations. This means that, as the temperature rises, the increase in energy is stronger even at intermediate densities, and the difference with its zero-temperature counterpart slowly reduces with density. A steeper decreasing behavior appears in PNM with respect to SNM, meaning that the energy at zero temperature in PNM is already a quite stiff quantity, leading to a smaller thermal energy.

\begin{figure*}
\centering
\includegraphics[width=0.45\textwidth]{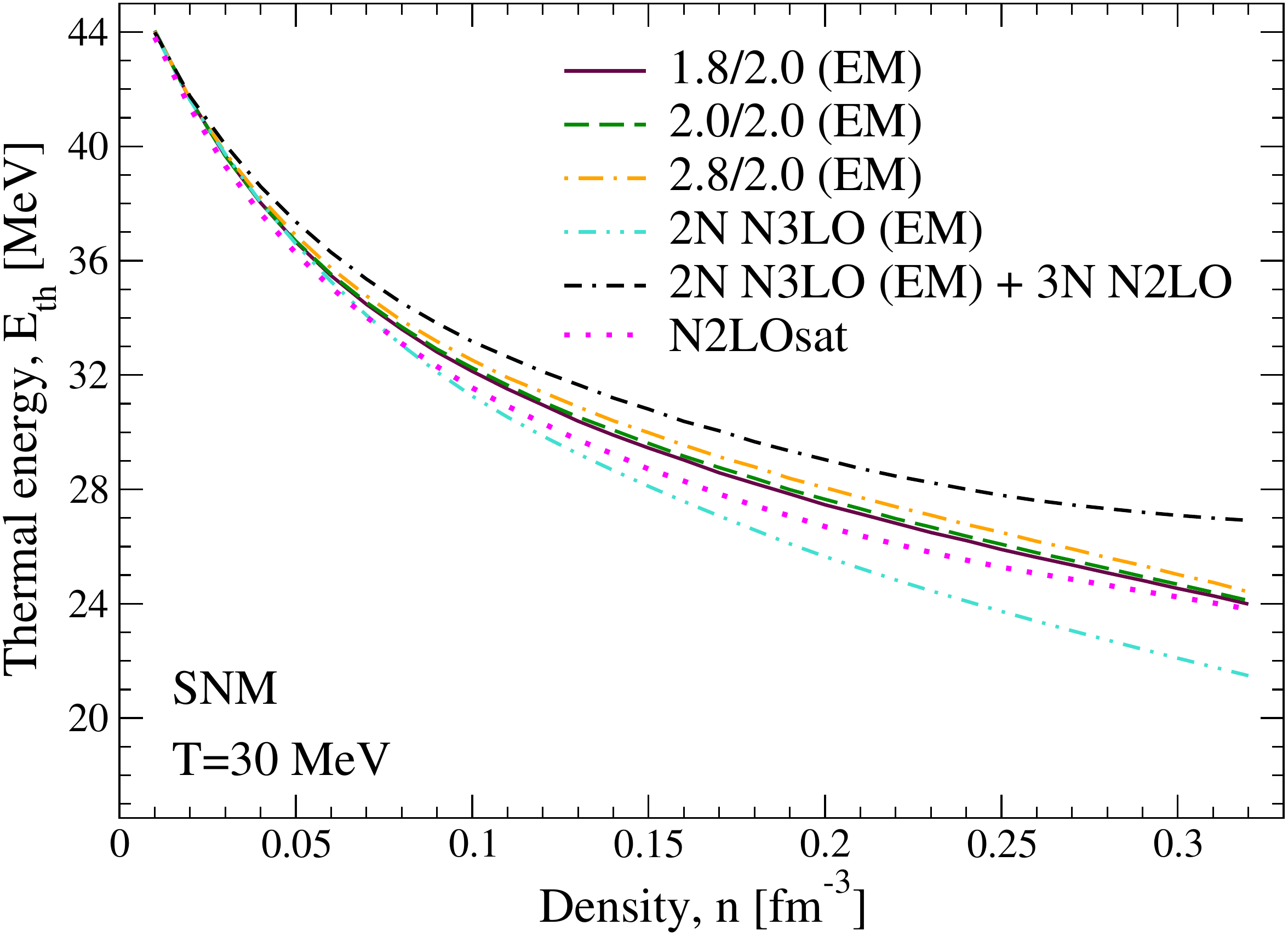}
\hspace*{5mm}
\includegraphics[width=0.45\textwidth]{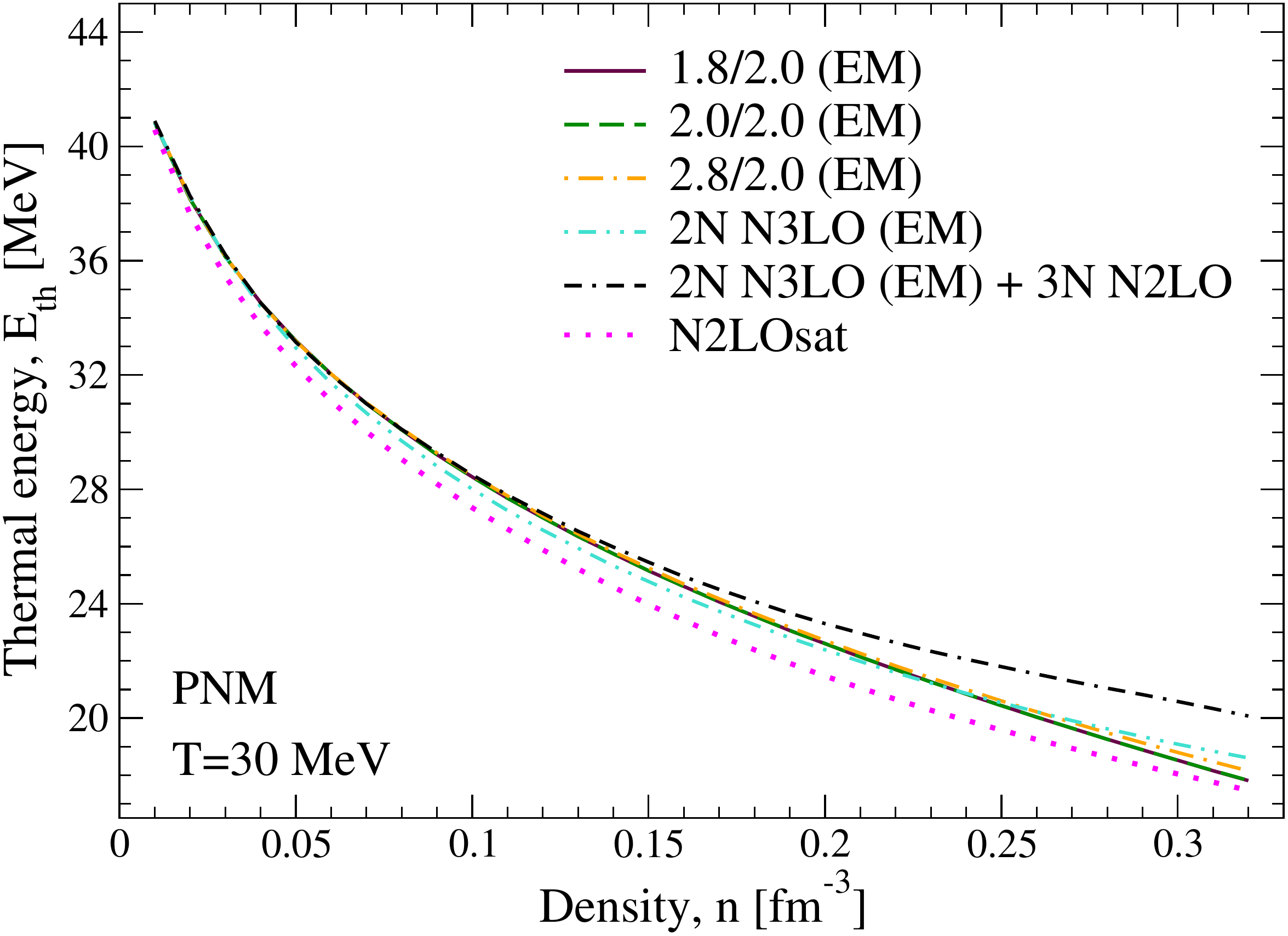}
\includegraphics[width=0.45\textwidth]{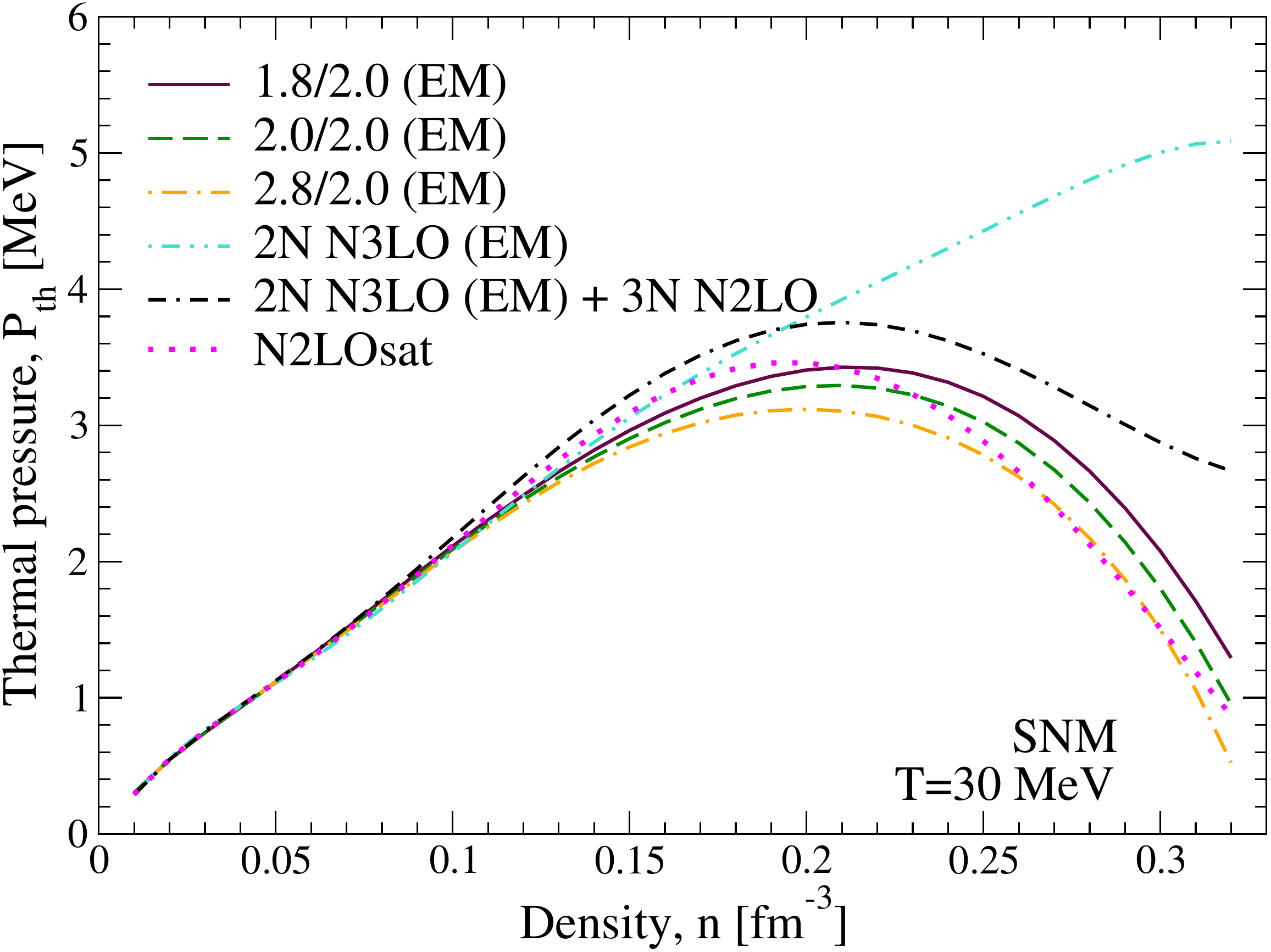}
\hspace*{5mm}
\includegraphics[width=0.45\textwidth]{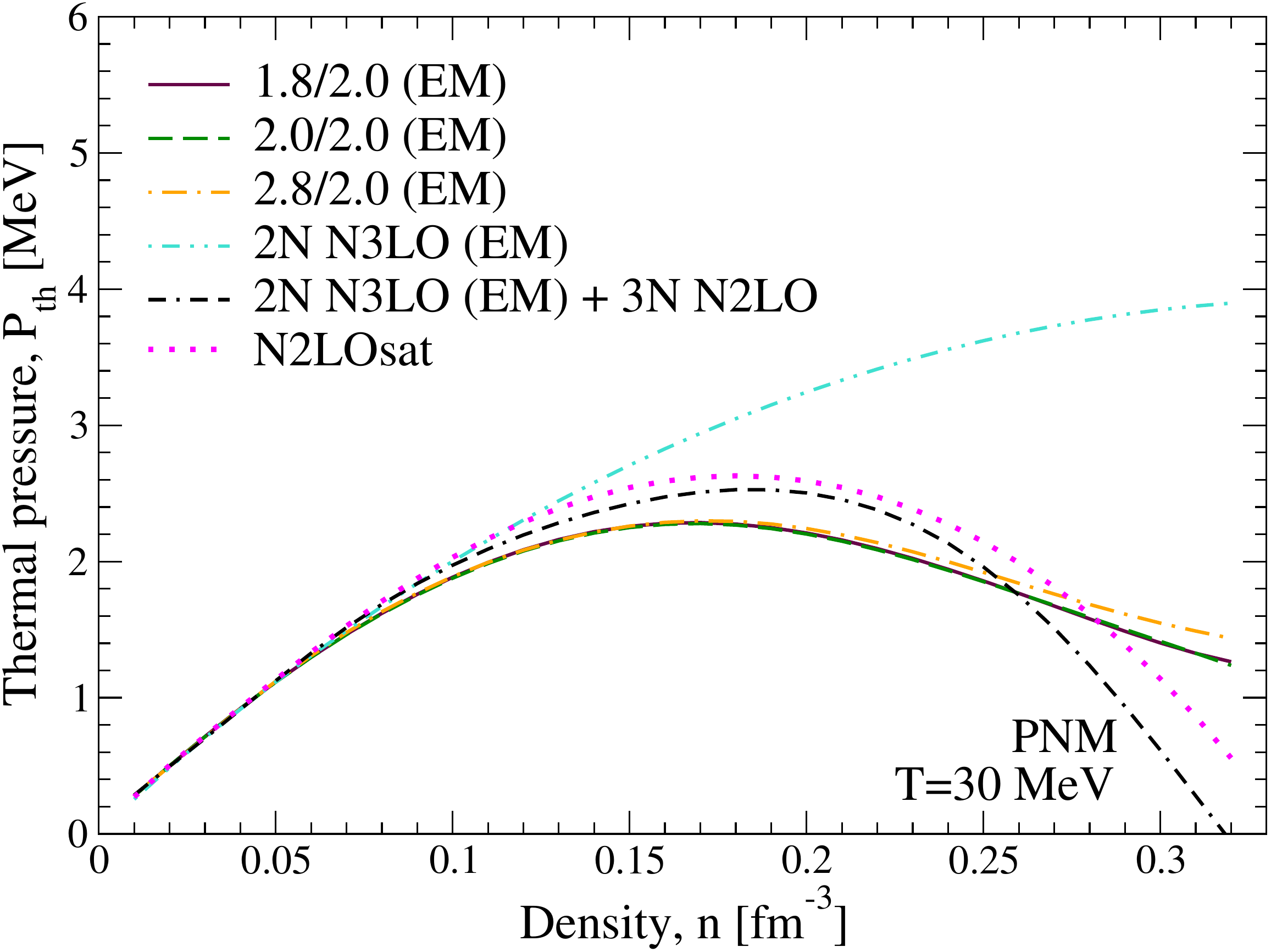}
\includegraphics[width=0.45\textwidth]{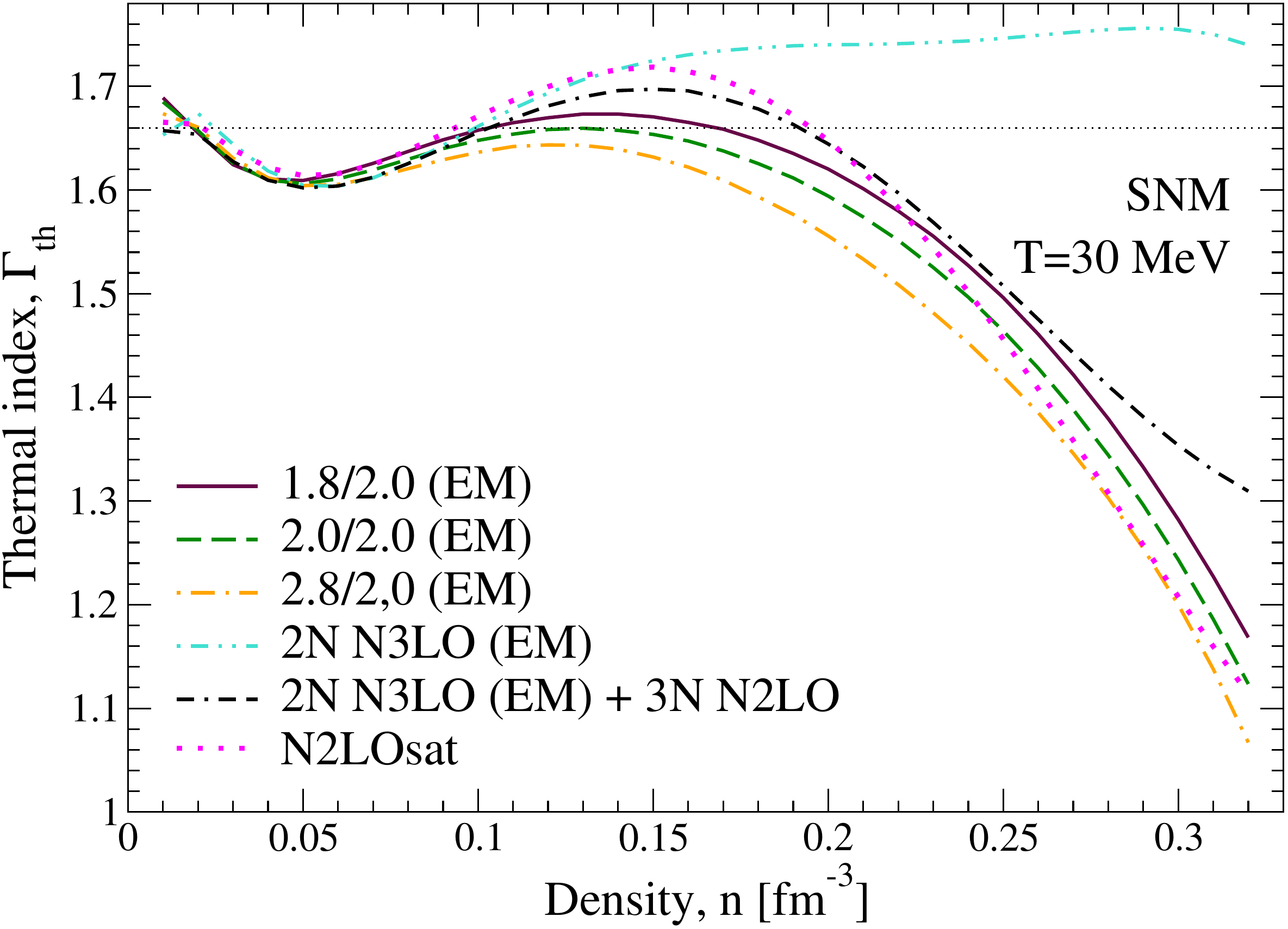}
\hspace*{5mm}
\includegraphics[width=0.45\textwidth]{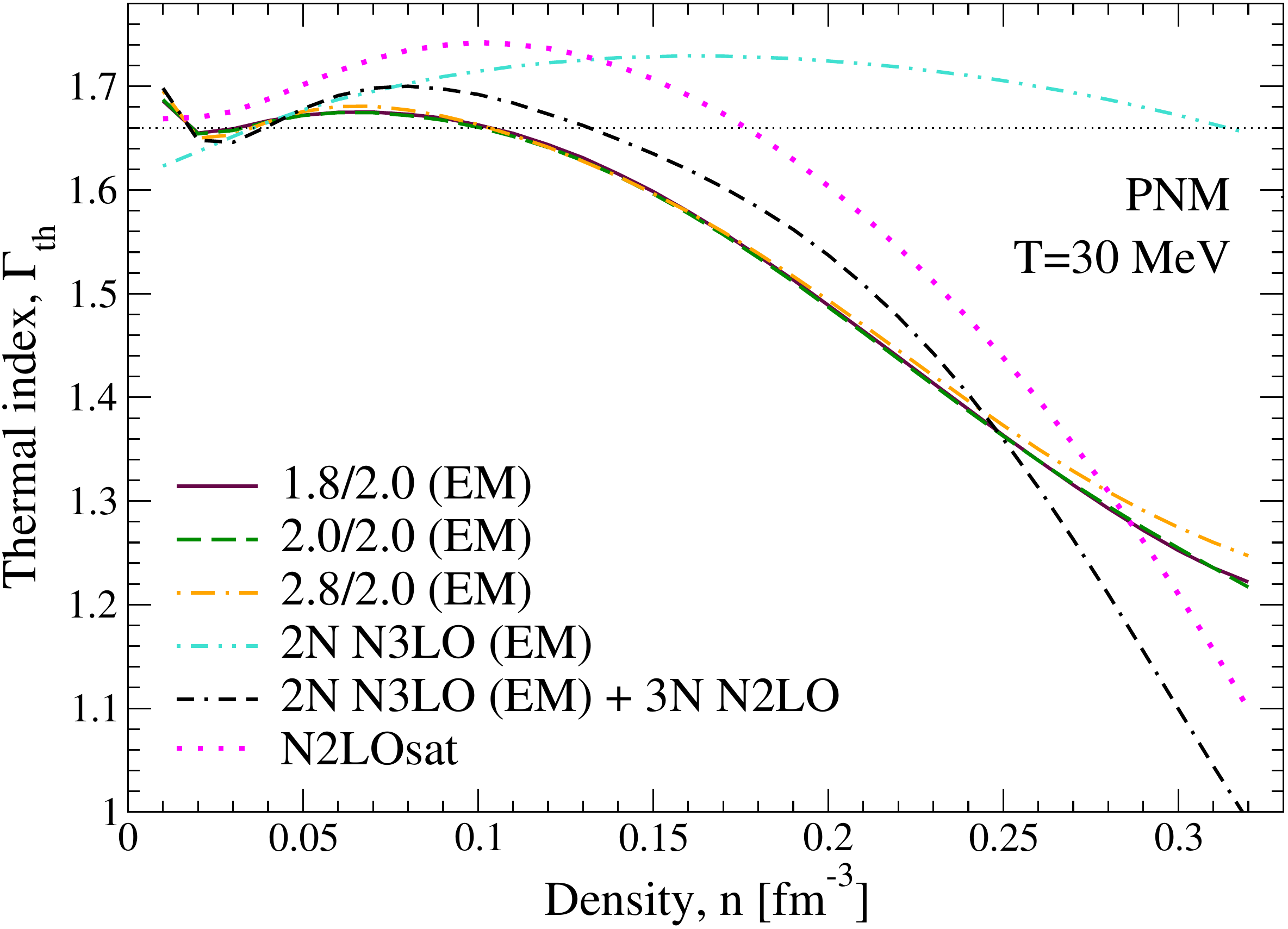}
\caption{Thermal energy (first row), thermal pressure (second row), and thermal index (third row) 
for $T=30$~MeV as function of density for SNM (left panels) and PNM (right panels),
using six different chiral two- and three-nucleon interactions (see text for details). Note that the $2N$
N3LO (EM) results are for two-nucleon interactions only.}
\label{fig:therm_comp}
\end{figure*}

We must point out that twice saturation density is close to the limit of validity of the chiral interactions considered. In fact, especially for PNM, one is already probing the range of the resolution scale $\lambda_{\rm SRG}=2.0$~fm$^{-1}$ for this particular Hamiltonian, where $\lambda_{\rm SRG}$ defines the similarity-renormalization-group resolution scale (see Ref.~\cite{Hebeler2011} for details). We explore the uncertainty in the high-density region (up to twice saturation density) by presenting calculations employing several chiral two- and three-nucleon interactions (see Fig.~\ref{fig:therm_comp} and following discussions), but note that the sensitivity to different interactions at high densities is probably only a lower bound on the uncertainty.

The thermal pressure $P_{\rm th}$, shown in the second row of Fig.~\ref{fig:therm_srg20}, presents a very different behavior from the thermal energy and, unlike $E_{\rm th}$, it is not only quantitatively different, but also qualitatively different between SNM and PNM. For SNM, the thermal pressure first increases, reaching a maximum around $0.20$~fm$^{-3}$ for all temperatures, and then keeps decreasing with increasing density. For PNM, the increase at low densities is much softer, with a maximum reached around saturation density, and the subsequent decrease being washed out with increasing temperature. From a quantitative point of view, the thermal pressure for SNM is bigger for intermediate densities, given that nuclear matter has negative zero-temperature pressure in this density region; the equivalent quantity for PNM is smaller, approaching nevertheless a higher value with respect to SNM at higher densities. This shows that, for increasing density, the pressure at finite temperature is stronger in PNM than SNM, and this strength rises with temperature. This is clearly visible in Fig.~\ref{fig:press}. Note that the high-density behavior of the thermal pressure is the one influencing the characteristics of the thermal index in this region.

In the last row of Fig.~\ref{fig:therm_srg20} we give the thermal index extracted using Eq.~\eqref{eq:thindex}. The qualitative behavior between SNM and PNM is similar, nevertheless quantitative differences appear. At low densities, the thermal index approaches the value of the adiabatic index for a nonrelativistic ideal gas with $\Gamma=5/3$; a small-dotted line is shown in the panels to guide the eye. The behavior observed for densities below $n=0.05$ fm$^{-3}$, especially for SNM at low temperatures, can be traced to the difficulty in fitting the free-energy in the low-density region (see Fig.~\ref{fig:free_ener}). For low temperatures, we observe a maximum for the thermal index which exceeds the value of 5/3. This maximum is more pronounced in SNM and appears around saturation density, while it is smaller for PNM, emerging in this case around half saturation density. However, as the temperature increases, this maximum smoothens due to a balance between the thermal pressure and the thermal energy. For higher densities, the behavior is dictated by the thermal pressure, as discussed above. In fact, while for SNM in Fig.~\ref{fig:therm_srg20} the thermal index shows a constant decrease, for PNM the decrease in $\Gamma_{\rm th}$ is levelled as the temperature increases, as it was observed for the respective thermal pressures. Note that for very high temperatures the system should behave as a relativistic gas, and relativistic effects should be taken into account. According to the limits imposed by Taub's inequality and to be consistent with relativistic kinetic theory, the adiabatic index should never exceed the value of 5/3, and should approach the value of 4/3 in the limits of high temperature~\cite{Mignone2007}. Indeed, even though we are performing a nonrelativistic calculation of nuclear matter, the thermal index seems like not exceeding 5/3 for higher temperatures and approaching the 4/3 value as density increases.  

We now turn to assessing an error estimate of the thermal effects employing several chiral Hamiltonians. We show in Fig.~\ref{fig:therm_comp} the thermal energy, thermal pressure, and thermal index for $T=30$~MeV for SNM and PNM. We use six different chiral interactions: the evolved chiral two- and three-nucleon interactions labeled 1.8/2.0 (EM), 2.0/2.0 (EM), and 2.8/2.0 (EM) are taken from Ref.~\cite{Hebeler2011}; the two-nucleon potential only from Entem and Machleidt~\cite{Entem2003} labeled $2N$ N3LO (EM); the $2N$ N3LO (EM) + $3N$ N2LO with a three-nucleon interaction fit to the triton beta decay from Ref.~\cite{Klos2016} (using the nonlocal 500/500 fit); and the N2LOsat two- and three-nucleon interaction from Ref.~\cite{Ekstroem2015}.

In the first row of Fig.~\ref{fig:therm_comp}, the thermal energy at $T=30$~MeV follows the behavior already described in Fig.~\ref{fig:therm_srg20}: the thermal energy decreases with increasing density. The spread provided by the use of different nuclear forces is more pronounced in SNM than in PNM. In SNM the extremes of the band are encompassed by the $2N$-only calculation, for the lower part, and the corresponding $2N+3N$ calculation, for the upper part. This shows the importance of three-nucleon interactions for the thermal energy. Chiral three-nucleon forces provide important repulsive contributions that increase with density, both in PNM~\cite{Hebeler2010} and SNM~\cite{Hebeler2011,Hagen:2013yba}. This is also the case at finite temperature and leads to larger thermal energies at higher densities. For PNM this effect is still visible, however the softer evolved interactions [1.8/2.0 (EM), 2.0/2.0 (EM), and 2.8/2.0 (EM)] result in thermal energies that are below the $2N$-only calculation, meaning that three-nucleon effects are weaker for these interactions at higher densities. It is interesting to note that this is also the case for the N2LOsat two- and three-nucleon interactions, which provide an even smaller thermal energy in PNM. 

In second row of Fig.~\ref{fig:therm_comp}, we present the thermal pressure obtained for the same chiral Hamiltonians. A striking characteristic is observed for both SNM and PNM: in the case of the $2N$-only calculation the thermal pressure is a growing quantity with density, unlike observed for all other cases. This hints at the fact that the zero temperature pressure is too soft without three-nucleon interactions, while the finite-temperature pressure stiffens caused by thermal components, providing a stronger thermal pressure. For all chiral two- and three-nucleon interactions, the thermal pressure decreases with density after having reached a maximum around $n=0.20$fm$^{-3}$. The decrease is a combined effect between how stiff the pressure is at zero temperature and how rapidly the finite-temperature pressure grows as density increases. 

Finally, the thermal index is shown in the last row of Fig.~\ref{fig:therm_comp}. As described above, an increase is observed at intermediate densities, for both SNM and PNM, followed by a rapid decrease as density grows. However, the case where only $2N$ interactions are included in the calculation presents a thermal index with a nearly density independent behavior, in comparison to the other cases. It is worth noting that for the unevolved potentials, namely $2N$ N3LO (EM) + $3N$ N2LO and N2LOsat, a higher maximum is reached; this is mostly caused by the stronger thermal pressure obtained with these chiral Hamiltonians. These interactions also show a steeper decrease in the PNM case, which is a direct consequence of the thermal pressure trend. This is caused by the fact that the zero-temperature pressure for these interactions is  stiffer due to stronger three-nucleon forces at high densities and so the thermal pressure is smaller at higher densities.

We conclude from this analysis that the thermal index behavior strongly depends on the inclusion of three-nucleon forces, on how important thermal effects are at intermediate densities for both energy and pressure, and on how stiff the pressure is at zero temperature in the high-density region. We observe a maximum value of $\approx \,$1.7, which varies slightly depending on the temperature (see last row of Fig.~\ref{fig:therm_srg20}) or nuclear forces considered (see last row of Fig.~\ref{fig:therm_comp}). As density increases, $\Gamma_{\rm th}$ reaches lower values down to $\approx \,$1 or around $1.3-1.4$ at higher temperatures. We never observe a value as high as $\Gamma_{\rm th} = 2$, which has been assessed as a reasonable value to simulate neutron-star mergers events~\cite{Bauswein2010,Rezzolla2016}. Note that a larger value for $\Gamma_{\rm th}$ refers to stiffer thermal effects, which leads to a smaller peak frequency in the gravitational-wave signal of the post-merger remnant and a longer time delay to black-hole collapse~\cite{Bauswein2010}. Finally, while we do not show uncertainty estimates from the many-body SCGF calculations, this is expected to be smaller than the spread in the different chiral Hamiltonians considered (see Ref.~\cite{Carbone2018,Carboneunpub}).

\subsection{Characterizing thermal effects through the nucleon effective mass}

\begin{figure*}
\centering
\includegraphics[width=0.45\textwidth]{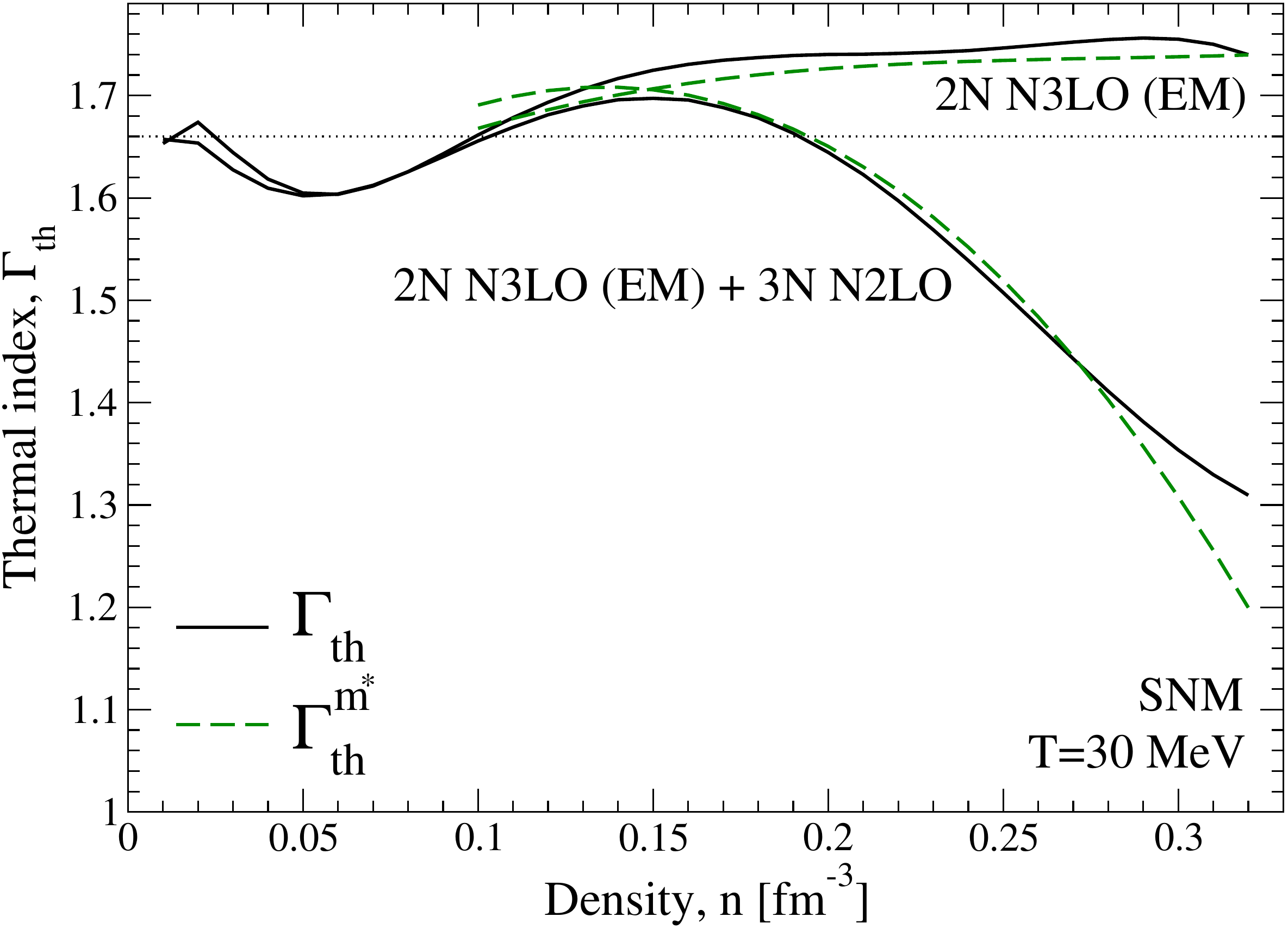}
\hspace*{5mm}
\includegraphics[width=0.45\textwidth]{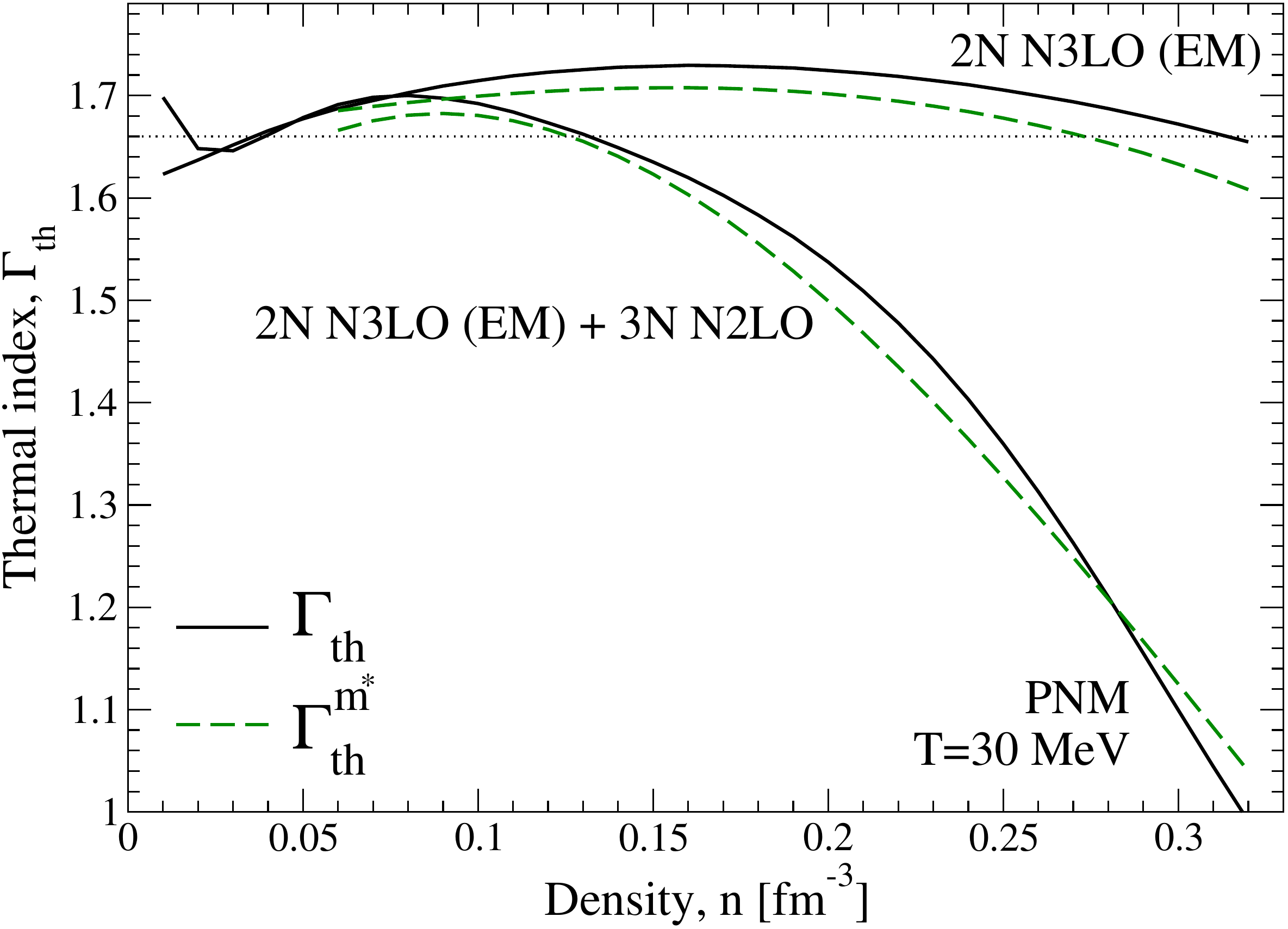}
\caption{Thermal index for $T=30$~MeV as function of density for SNM (left) and PNM (right).
The black solid lines are extractions of the thermal index from the thermal energy and thermal
pressure using Eq.~\eqref{eq:thindex} while the green dashed lines are for 
$\Gamma_{\rm th}$ based on the density dependence of the effective nucleon mass, Eq.~\eqref{eq:thindex_mass}. Results are shown for the $2N$ N3LO (EM) and $2N$ N3LO (EM) 
+ $3N$ N2LO chiral interactions.}
\label{fig:th_effmass}
\end{figure*}

\begin{figure*}
\centering    
\includegraphics[width=0.45\textwidth]{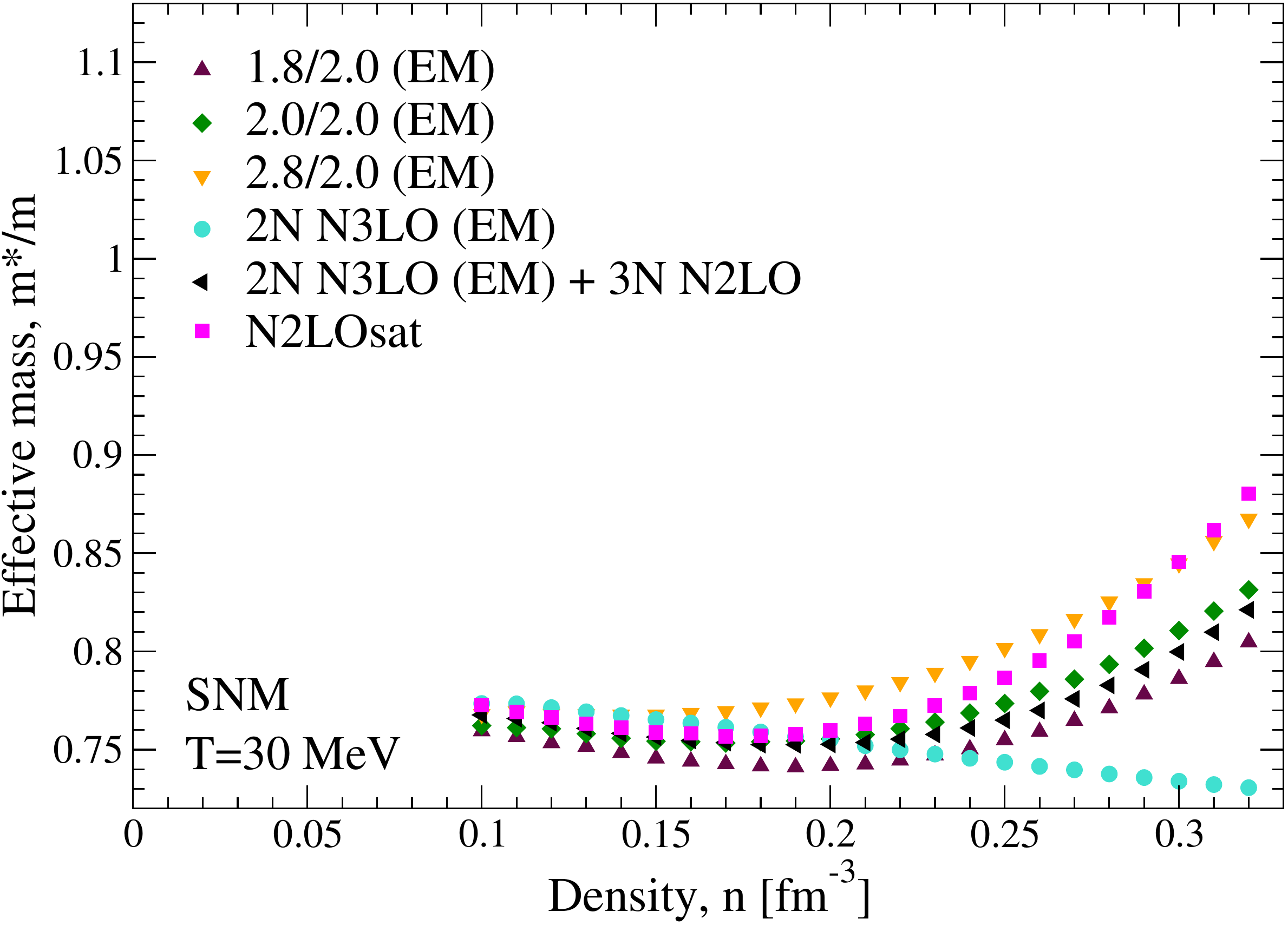}
\hspace*{5mm}
\includegraphics[width=0.45\textwidth]{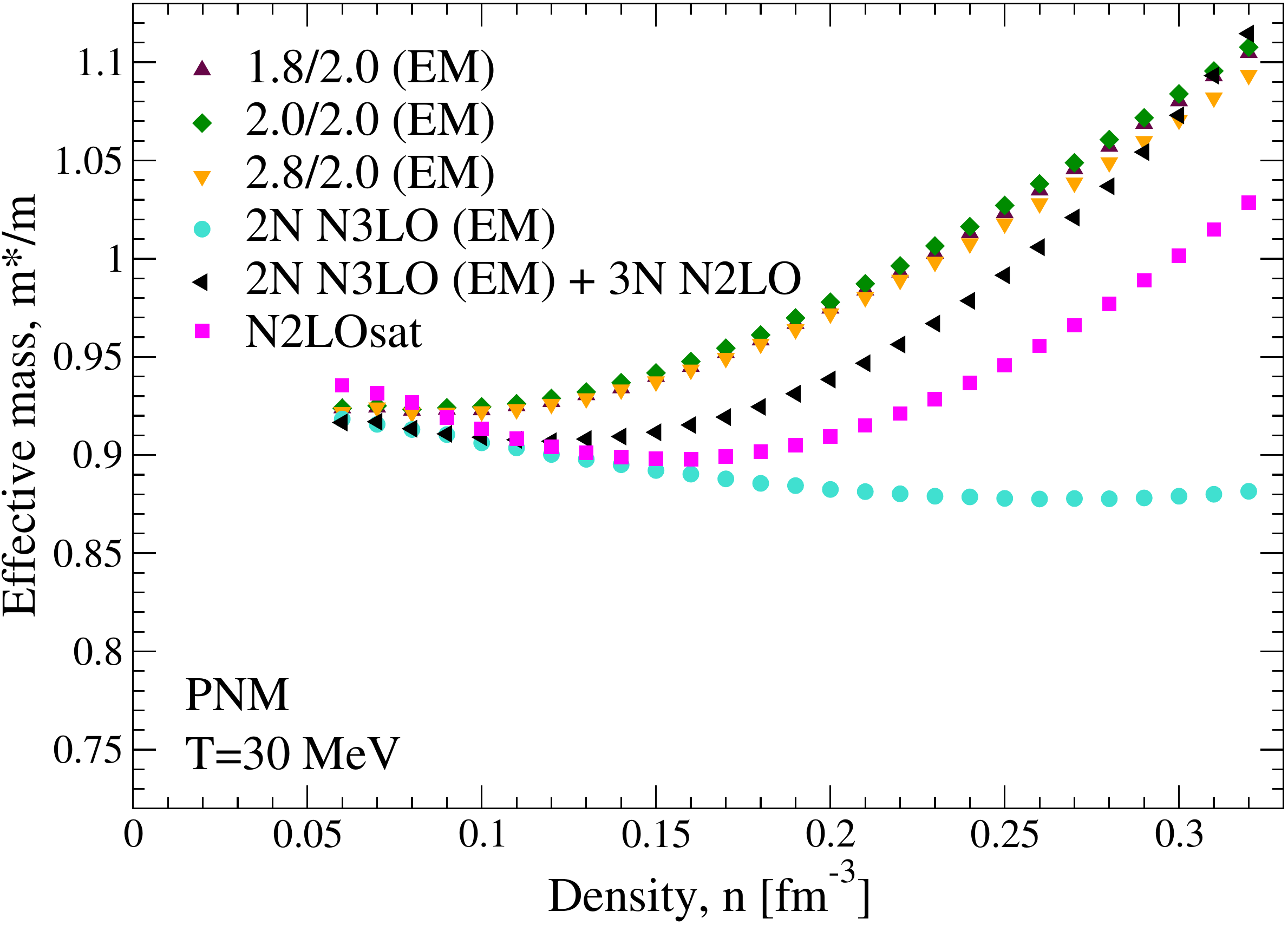}
\caption{Effective mass for $T=30$~MeV as function of density for SNM (left) and PNM (right),
extracted from the single-particle energy at the chemical potential [see Eq.~\eqref{eq:effmass_mu}] 
using six different chiral two- and three-nucleon interactions.}
\label{fig:effmass_all}
\end{figure*}

\begin{figure*}
\centering
\includegraphics[width=0.45\textwidth]{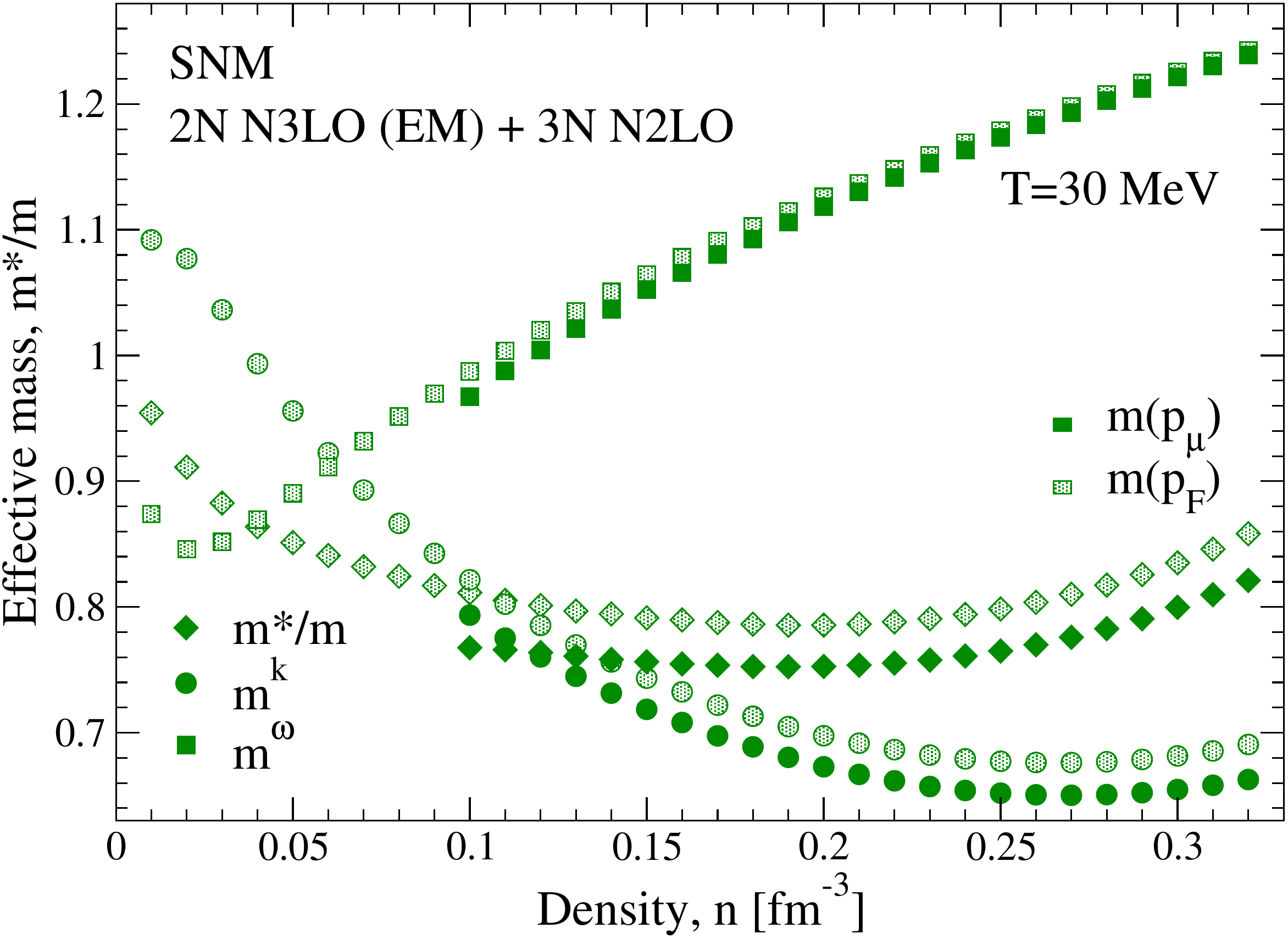}
\hspace*{5mm}
\includegraphics[width=0.45\textwidth]{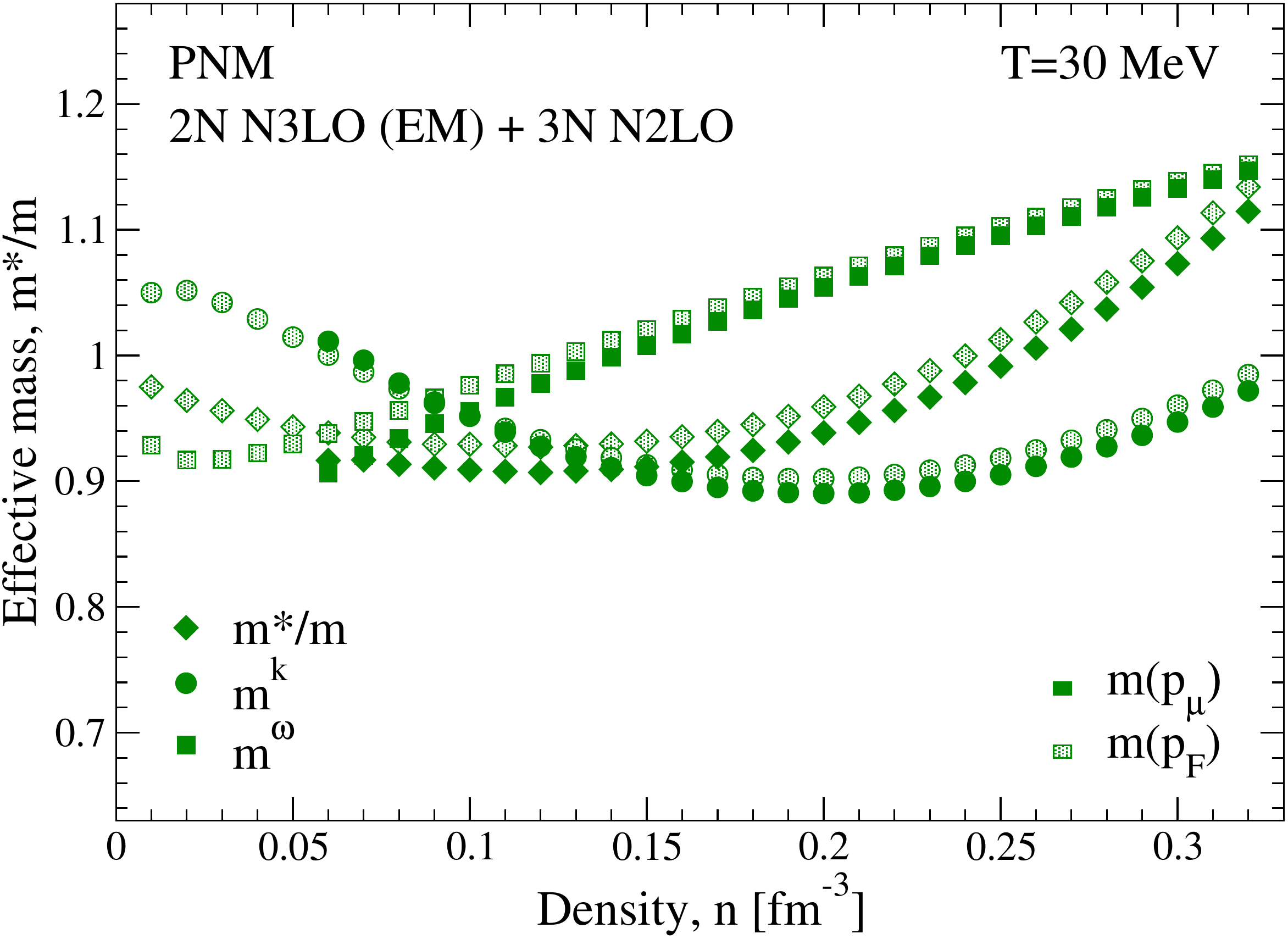}
\includegraphics[width=0.45\textwidth]{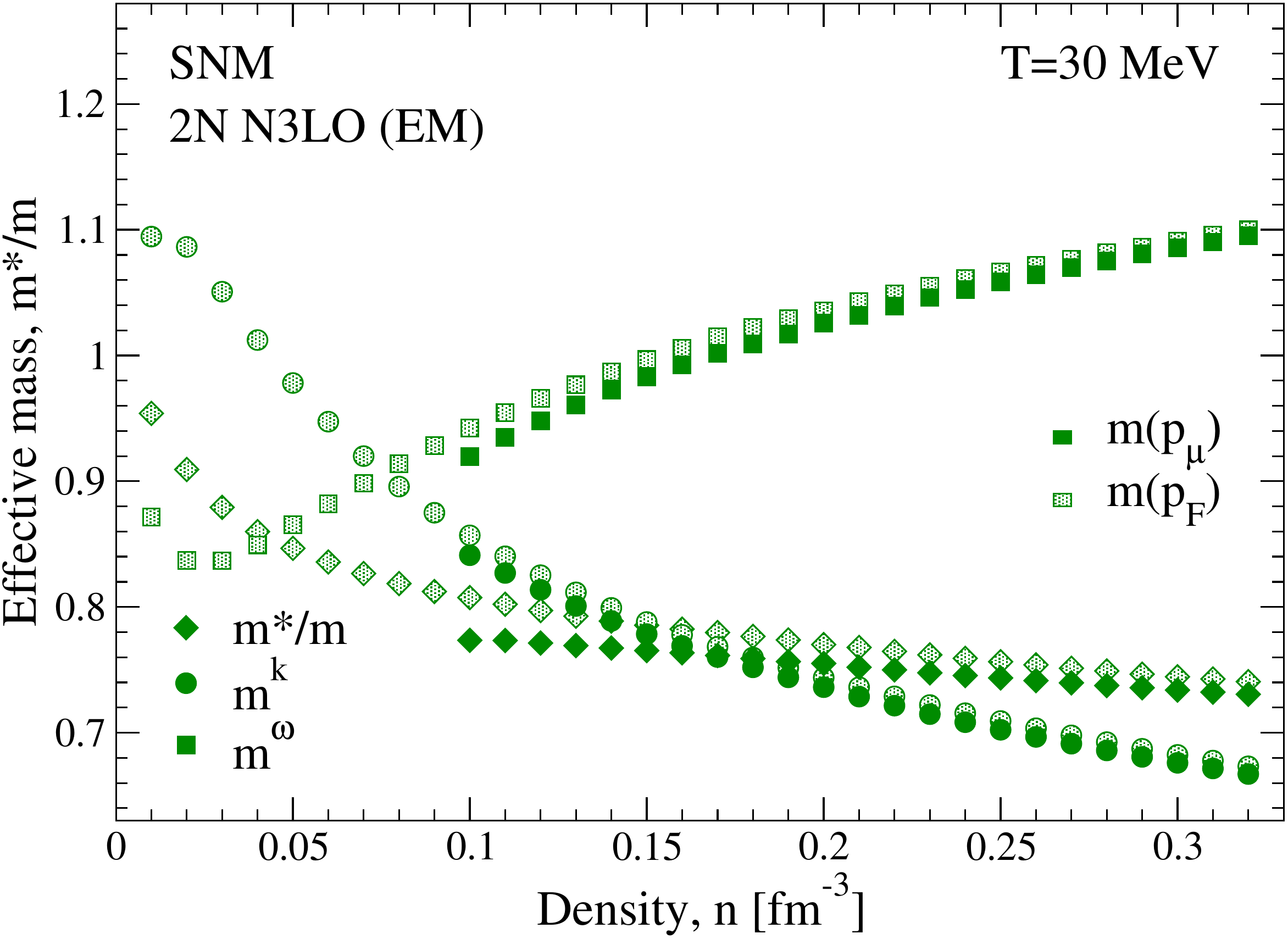}
\hspace*{5mm}
\includegraphics[width=0.45\textwidth]{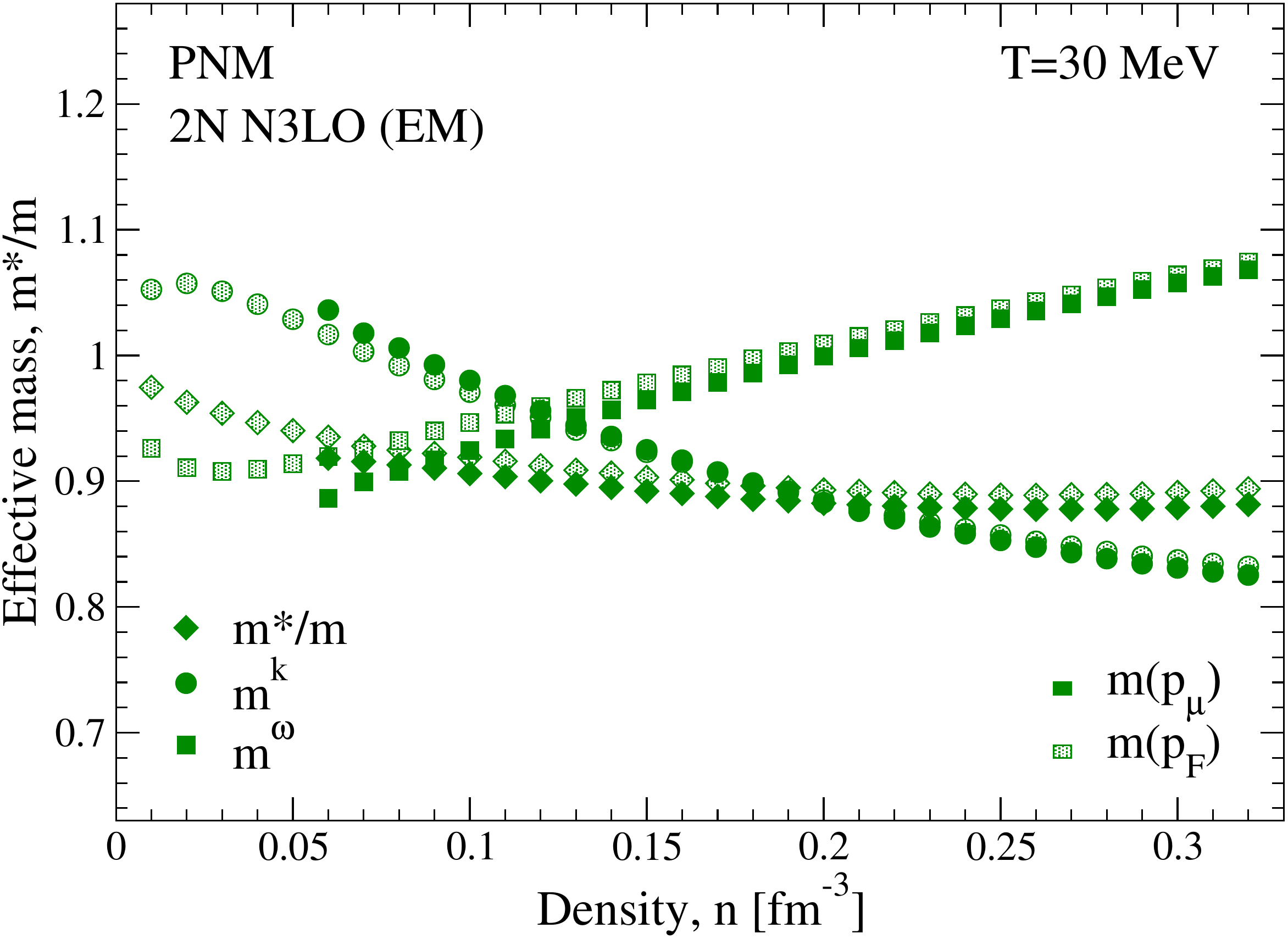}
\caption{Effective mass for $T=30$~MeV as function of density for SNM (left panels) and PNM 
(right panels) using the $2N$ N3LO (EM) + $3N$ N2LO (upper row) and $2N$ N3LO (EM) (lower row) 
chiral interactions. The full $m^*$ as well as $m^{\omega}$ and $m^k$ are shown, extracted
from the single-particle energy at the chemical potential, $m(p_{\mu})$, or at the Fermi 
momentum, $m(p_{\rm F})$.}
\label{fig:effmass}
\end{figure*}

In this section, we show how knowledge of the nucleon effective mass sheds light on the behavior of the thermal index. To this end, we explore a functional form, which is based on Fermi liquid theory of a weakly interacting nonrelativistic system of quasiparticles with a density-dependent effective mass $m^*$. In fact, assuming an ideal gas of nucleons with density-dependent $m^*(n)$ yields for the thermal index the following form (see, e.g., Ref.~\cite{Constantinou2015}):
\begin{equation}
\Gamma_{\rm th}^{m^*}=\frac{5}{3}-\frac{n}{m^*}\frac{\partial m^*}{\partial n}\,,
\label{eq:thindex_mass}
\end{equation}
which in the limit of zero density approaches the nonrelativistic ideal gas index 5/3. 

In Fig.~\ref{fig:th_effmass}, we compare the thermal index for $T=30$~MeV extracted from the thermal energy and thermal pressure using Eq.~\eqref{eq:thindex} to $\Gamma_{\rm th}$ based on the density dependence of the effective nucleon mass, Eq.~\eqref{eq:thindex_mass}. The results are shown for the $2N$ N3LO (EM) and $2N$ N3LO (EM) + $3N$ N2LO chiral interactions for SNM and PNM. Similar results are found for the other temperature and Hamiltonians studied before. It is remarkable how well the functional form given by Eq.~\eqref{eq:thindex_mass} captures the behavior of the thermal index, with only small quantitative differences. This demonstrates that by far the dominant thermal effects are determined by the density-dependent effective mass of the nucleons. We note that the reason why the dashed lines in Fig.~\ref{fig:th_effmass} do not capture the low-density behavior can be traced to how we determine the effective mass at finite temperature. In fact, we calculate the effective mass from the single-particle spectrum where the single-particle energy equals the chemical potential (see Ref.~~\cite{BALi2018} for a recent review on nucleon effective masses):
\begin{equation}
\frac{m^*}{m}=\frac{p}{m}\left(\frac{\partial\varepsilon(p)}{\partial p}\right)^{-1}\bigg|_{\varepsilon(p)=\mu,p=p_\mu}\,,
\label{eq:effmass_mu}
\end{equation}
where $p_\mu$ defines the momentum where the single-particle energy $\varepsilon(p)$ equals the chemical potential. In the low-density regime, especially at higher temperatures, the chemical potential becomes very negative and no solution is encountered for the equality $\varepsilon(p)=\mu$, which is why the dashed lines stop at certain low-density values, for both SNM and PNM.

In Fig.~\ref{fig:effmass_all} we plot the effective mass for $T=30$~MeV as determined by Eq.~\eqref{eq:effmass_mu}, exploring the different chiral Hamiltonians for SNM and PNM. We find a qualitatively similar behavior for SNM and PNM. The effective mass first decreases and then increases with density, except for the $2N$-only calculation, where the effective mass keeps decreasing for SNM, while it slightly grows for PNM, when reaching twice saturation density. It is worth noting that for PNM the effective mass is much closer to 1, and reaches values above 1 as density increases. The inclusion of three-nucleon forces provides important repulsive contributions to the self-energy, which was already discussed for the thermal index in Fig,~\ref{fig:therm_comp}, and can be clearly seen also for the effective mass in Fig.~\ref{fig:effmass_all}. We have checked that at low temperatures our effective masses compare reasonably well with zero temperature results from Refs.~\cite{Hebe09enerfunc,Dris14asymmat,Well14nmtherm,Dris16gap,Burac2019}. Finally, we note that at densities where the three-nucleon contributions are small, the results for the effective mass are consistent with previous studies~\cite{Hebe09enerfunc,Well14nmtherm}.

To understand the behavior of the effective mass, we present in Fig.~\ref{fig:effmass} its different momentum- and energy-dependent contributions, the so called $k$- and $\omega$-mass. We focus on the same $T=30$~MeV temperature and the same chiral two- and three-nucleon interactions studied in Fig.~\ref{fig:th_effmass}. Considering the different contributions the full effective mass of Eq.~\eqref{eq:effmass_mu} can be written as a product of the $k$ and $\omega$ mass, $m^{k}$ and $m^{\omega}$,
\begin{equation}
\frac{m^*}{m}=m^{\omega}m^k\,,
\label{eq:effmass_full}
\end{equation}
where $m^{\omega}$ is given by the energy derivative of the real part of the self-energy,
\begin{equation}
m^{\omega}=1-\frac{\partial{\rm Re}\Sigma(p,\omega)}{\partial\omega}\,,
\label{mw}
\end{equation}
and $m^k$ by the momentum derivative,
\begin{equation}
m^k=\left(1+\frac{m}{p}\frac{\partial{\rm Re}\Sigma(p,\omega)}{\partial p}\right)^{-1}\,.
\label{mk}
\end{equation}
${\rm Re}\Sigma(p,\omega)$ describes the real part of the self-energy which we calculate within the SCGF approach. Note that at the Hartree-Fock level or in the mean-field approximation, the effective mass would be given by $k$ mass only. Figure~\ref{fig:effmass} shows clearly that $m^{\omega}$ from energy-dependent correlations causes the increase of the effective mass with density. However, although $m^{\omega}$ is larger in SNM than in PNM, it is nevertheless balanced by a smaller value for $m^k$ in the former case, producing a smaller total effective mass. When three-nucleon forces are not included (in the second row of Fig.~\ref{fig:effmass}), the effects of $m^{\omega}$ are not as strong, both in SNM and PNM, and a combined effect with a decreasing $m^k$ as function of density is the main cause of the nearly density independent effective mass at higher densities, as observed in Fig.~\ref{fig:effmass_all}. In all panels of Fig.~\ref{fig:effmass} we also show for comparison the effective mass calculated at the Fermi momentum determined from the density; we see that differences are larger at low densities, where thermal effects are stronger and $p_\mu$ differs more from $p_{\rm F}$.

We thus find important contributions to the effective mass beyond the Hartree-Fock level or mean-field approximation, and that a full description of the effective mass, as given by Eq.~\eqref{eq:effmass_full}, is  important to describe thermal effects for the nuclear EOS and to reproduce the behavior of the thermal index.

\section{Conclusions and outlook}
\label{sec:conclusions}

We have presented first {\it ab initio} SCGF calculations of thermal effects on the nuclear EOS using different chiral two- and three-nucleon interactions. In particular, we analyzed for SNM and PNM the thermal energy and thermal pressure, from which we accessed the behavior of the thermal index used widely in astrophysical simulations. Our calculations show how a density-dependent $\Gamma_{\rm th}$, which is, e.g., based on ideal-gas thermal contributions, does not capture the thermal effects based on {\it ab initio} calculations.

We have provided uncertainty estimates employing different chiral Hamiltonians. Overall the thermal index was found to vary between about 1-1.7 according to changes in density or temperature, and including the nuclear physics uncertainties. The behavior of the thermal index is strikingly affected by the inclusion of three-nucleon forces in our calculations; furthermore the stiffness of the pressure at zero temperature influences the decrease of $\Gamma_{\rm th}$ at high densities. Our results clearly point to smaller values for the thermal index, compared to the range of $\Gamma_{\rm th}$ up to 2 used in astrophysical simulations. Such low values are also expected to have a very interesting impact on core-collapse supernova simulations, based on the recent work of Ref.~\cite{Yasin2018}.

We have also explored a functional form for $\Gamma_{\rm th}$ based on the density dependence of the nucleon effective mass, which captures the behavior of the thermal index remarkably well. In particular, this shows that a calculation of the effective mass, beyond the Hartree-Fock level or mean-field approximation, is necessary to capture these thermal effects. This work is a first step towards a more comprehensive analysis of thermal effects in the nuclear EOS and a full finite-temperature description based on realistic nuclear interactions. This will help shed light on the dependence of the threshold mass to prompt collapse in neutron star mergers as a function of the maximum compactness~\cite{Bauswein2013,Koppel2019}, and also establish more realistic lower bounds for the binary tidal deformability extracted from electromagnetic counterparts~\cite{Kiuchi2019}. Future improvements will concentrate on asymmetric matter in beta-equilibrium, as well as extensions to explore higher densities, relevant for astrophysical applications. 

\begin{acknowledgments}
We thank A.~Arcones, K.~Hebeler, L.~Rezzolla, S.~Sch\"afer, C.~Wellenhofer, and H.~Yasin for useful discussions. This work was supported in part by the Deutsche Forschungsgemeinschaft (DFG, German Research Foundation) -- Projektnummer 279384907 -- SFB 1245 and by ``PHAROS" COST Action CA16214. Calculations for this research were conducted on the Lichtenberg high-performance computer of the TU Darmstadt.
\end{acknowledgments}

\bibliography{biblio.bib}

\end{document}